\documentclass[fleqn,usenatbib]{mnras}

\usepackage[T1]{fontenc}

\DeclareRobustCommand{\VAN}[3]{#2}
\let\VANthebibliography\thebibliography
\def\thebibliography{\DeclareRobustCommand{\VAN}[3]{##3}\VANthebibliography}

\def\equationautorefname#1#2\null{(#2)}

\usepackage{graphicx}	 % Including figure files
\usepackage{amsmath}	 % Advanced maths commands
\usepackage{amssymb}	 % Extra maths symbols
\usepackage{mathtools}   % multi-lines for equation
\usepackage{CJK}         % Chinese name
\usepackage{xcolor}      % comments color
\usepackage{scalerel}
\usepackage{tikz}
\usetikzlibrary{svg.path}
\definecolor{orcidlogocol}{HTML}{A6CE39}
\tikzset{
  orcidlogo/.pic={
    \fill[orcidlogocol] svg{M256,128c0,70.7-57.3,128-128,128C57.3,256,0,198.7,0,128C0,57.3,57.3,0,128,0C198.7,0,256,57.3,256,128z};
    \fill[white] svg{M86.3,186.2H70.9V79.1h15.4v48.4V186.2z}
                 svg{M108.9,79.1h41.6c39.6,0,57,28.3,57,53.6c0,27.5-21.5,53.6-56.8,53.6h-41.8V79.1z M124.3,172.4h24.5c34.9,0,42.9-26.5,42.9-39.7c0-21.5-13.7-39.7-43.7-39.7h-23.7V172.4z}
                 svg{M88.7,56.8c0,5.5-4.5,10.1-10.1,10.1c-5.6,0-10.1-4.6-10.1-10.1c0-5.6,4.5-10.1,10.1-10.1C84.2,46.7,88.7,51.3,88.7,56.8z};
  }
}

\usepackage{newtxtext,newtxmath}

\newcommand{\lz}[1]{#1}
\newcommand{\ob}[1]{#1}

\newcommand{\eqnref}[1]{equation~\autoref{#1}}
\newcommand{\BSmodel}{1D standard model}
\newcommand\orcidicon[1]{\href{https://orcid.org/#1}{\mbox{\scalerel*{
\begin{tikzpicture}[yscale=-1,transform shape]
\pic{orcidlogo};
\end{tikzpicture}s
}{|}}}}%%%%%%%%%%%%%%%%%%%%%%%%%%%%%%%%%%%%%%%%%%%%%%%%%%

\title[Cartesian Accretion Column]{Radiative \lz{Relativistic Magnetohydrodynamic} Simulations of Neutron Star Column Accretion in Cartesian Geometry}

\author[L. Zhang et al.]{
Lizhong Zhang (张力中) \orcidicon{0000-0003-0232-0879},$^{1}$\thanks{E-mail: lizhong@physics.ucsb.edu}
Omer Blaes \orcidicon{0000-0002-8082-4573},$^{1}$
Yan-Fei Jiang (姜燕飞) \orcidicon{0000-0002-2624-3399}$^{2}$
\\
$^{1}$Department of Physics, University of California, Santa Barbara, CA 93106, USA\\
$^{2}$Center for Computational Astrophysics, Flatiron Institute, New York, NY 10010, USA\\
}

\date{Accepted 2022 June 24. Received 2022 May 26; in original form 2022 January 7}

\pubyear{2021}

\begin{document}
\begin{CJK*}{UTF8}{gbsn}
\label{firstpage}
\pagerange{\pageref{firstpage}--\pageref{lastpage}}
\maketitle

\begin{abstract}
High luminosity accretion onto a strongly magnetized neutron star results in a radiation pressure \lz{dominated}, magnetically confined accretion column.  \lz{We investigate the dynamics of these columns} using two-dimensional radiative \lz{relativistic} magnetohydrodynamic simulations, restricting consideration to modest accretion rates where the height of the column is low enough that Cartesian geometry can be employed.  The column structure is dynamically maintained through high-frequency oscillations of the accretion shock at $\simeq 10-25$~kHz.  \lz{These oscillations arise because it is necessary to redistribute the power released at the accretion shock through bulk vertical motions, both to balance the cooling and to provide vertical pressure support against gravity.  Sideways cooling always dominates the loss of internal energy.  In addition to the vertical oscillations, photon bubbles form in our simulations and add additional spatial complexity to the column structure.  They are not themselves responsible for the oscillations, and they do not appear to affect the oscillation period.  However, they} enhance the vertical transport of radiation and increase the oscillation amplitude in luminosity.  The time-averaged column structure in our simulations resembles the trends in standard 1D stationary models, the main difference being that the time-averaged height of the shock front is lower because of the higher cooling efficiency of the \lz{2D column shape}. 
\end{abstract}


%%%%%%%%%%%%%%%%%%%%%%%%%%%%%%%%%%%%%%%%%%%%%%%%%%

%%%%%%%%%%%%%%%%% BODY OF PAPER %%%%%%%%%%%%%%%%%%

\section{Introduction}

In marked contrast to black holes, the accretion of matter onto a neutron star can be guided by a sufficiently strong stellar magnetic field to form surface hot spots or magnetically confined accretion columns.  If these spots or columns are not axisymmetrically distributed around the neutron star spin axis, the result is an accretion-powered X-ray pulsar.  Accreting X-ray pulsars, most of which are in high mass X-ray binaries, have been known since the early days of X-ray astronomy (see e.g. \citealt{Caballero_Wilms2012} for a review).  Accreting millisecond pulsars in low mass X-ray binaries were discovered much later \citet{Wijnands_vanderKlis1998}.  Beginning with the discovery by \citet{2014Natur.514..202B}, some ultraluminous X-ray sources are now known to exhibit coherent pulsations and are therefore clearly accreting, magnetized, \lz{rotating} neutron stars.

The theoretical structure of magnetically confined accretion flows on neutron stars was first elucidated by \citet{Inoue1975} and \citet{1975A&A....42..311B, 1976MNRAS.175..395B}.  When the accretion rate is low, a `pencil-beam' emission pattern of radiation develops due to the anisotropic effects of magnetic opacity above the hot spot on the surface of the star. When the accretion rate is high enough, a radiating shock forms above the surface of the neutron star and divides the column structure into two zones: 1. a free-fall zone above the shock, where the kinetic energy of the incoming material is converted into radiation at the shock front; and 2. a radiation pressure dominated sinking zone below the shock, where gravitational energy is slowly released.  In this case, a `fan-beam' emission pattern emerges as the accretion column radiates through its sides.

Approximate one-dimensional profiles of the column (hereafter called the \BSmodel) can be solved analytically assuming a horizontal one zone finite difference and a stationary column structure.  However, \citet{1989ESASP.296...89K} first showed by numerical simulation that accretion columns \lz{exhibited time-dependent structures which they associated with so-called ``photon bubbles'',} suggesting that \lz{accretion columns} cannot be stationary structures.  Since that discovery, the photon bubble instability has been explored extensively both analytically \citep{1992ApJ...388..561A, 1998MNRAS.297..929G, 2001ApJ...551..897B, 2003ApJ...596..509B, 2006ApJ...636..995B} and numerically \citep{1997ApJ...478..663H, 2005ApJ...624..267T, Zhang2021}. In accretion columns with high accretion rates, the sinking zone is radiation pressure dominated, and (to quote \citealt{1998MNRAS.297..929G}) this results in an inherently fragile hydrostatic equilibrium because the thermal pressure support is then independent of gas density. Vertical perturbations in density are therefore dynamically neutral (zero frequency entropy modes), but when horizontal gradients exist, the resulting heat flow can drive unstable phase relationships between density and radiation pressure \lz{perturbations}.  This is the photon bubble instability in the so-called slow-diffusion regime (\citealt{1992ApJ...388..561A}, Appendix A3 of \citealt{Zhang2021}).  The photon bubble instability can disrupt the equilibrium of the accretion column and therefore understanding the actual structure of such columns requires numerical simulations. 

In a previous paper \citep{Zhang2021}, we coupled the radiation transport algorithms \citep{2014ApJS..213....7J, 2021ApJS..253...49J} to the special relativistic magnetohydrodynamic (MHD) module \citep{2011ApJS..193....6B} \lz{in} the \textsc{Athena++} code framework \citep{STO20} to simulate the development of the photon bubble instability in static, magnetized neutron star atmospheres in Cartesian geometry.  The instability inevitably caused the atmosphere to collapse because the density inhomogeneities associated with the instability enhanced photon escape, resulting in a loss of radiation pressure support.  Similar behavior was earlier observed in simulations by \citet{1997ApJ...478..663H}, who used flux-limited diffusion and restricted the motion of the fluid along infinitely rigid magnetic field lines.  Our algorithm incorporates dynamical magnetic fields as well as a full solution to the time-dependent, frequency-integrated \lz{and angle-dependent} radiative transfer equation.  Our results were in excellent agreement with predictions of the linear growth of the instability \citep{1992ApJ...388..561A,1998MNRAS.297..929G, Zhang2021}.  In particular, the length scale of the fastest growing unstable modes \lz{decreases} with increasing numerical resolution, until a scale is reached where radiation viscosity starts to damp the modes.  Because we solve the angle-dependent transfer equation, radiation viscosity is automatically incorporated in our numerical treatment.

Since the pioneering simulations of unstable accretion columns by \citet{1989ESASP.296...89K} and \citet{1996ApJ...457L..85K}, there has been little further work done on this problem (though see \citealt{2020PASJ...72...15K}).  With the successful validation of our code in \citet{Zhang2021}, we here present simulations of the dynamics of vertically stratified neutron star accretion columns where mass is supplied from the top of the simulation domain.  We incorporate the sideways exterior of the accretion column within our simulation domain, thereby allowing radiation to escape from the sides \citep{Inoue1975,1976MNRAS.175..395B} as well as the top, and naturally incorporate the magnetic tension that confines the column against sideways radiation pressure.  Because we incorporate accretion by continuing to supply mass from the top, the accretion column does not collapse under photon bubble instability, but instead exhibits highly \lz{variable} behavior. 

Since the photon bubble instability is resolution-dependent, we continue to use Cartesian grids so that the photon bubble dynamics in the accretion column is resolved in a spatially unbiased manner. In any other geometry, the simulation grids would not be uniform and thus the instability would develop faster in the better resolved regions.  In this paper, we set up the simulations of Cartesian accretion columns with gas in 2D and radiation in 3D, which can be physically interpreted as a vertical slice of a narrow wall-shaped column. 

\lz{While photon bubbles are the historical motivating factor behind our simulations, we find in fact that they are not the dominant factor in the column dynamics.  Instead, the columns exhibit vertical oscillations that fundamentally originate from a mismatch between heating and cooling time scales.  We also observe photon bubbles, that further complicate the structure of the accretion column, but they do not fundamentally alter this basic oscillatory phenomenon.}

In \hyperref[sec:numerical_method]{Section 2}, we review the equations of the conservation laws in gas and radiation, the numerical setup of the simulation domain, initial conditions, boundary conditions, \lz{numerical issues associated with variable inversion, and simulation parameters}. In \hyperref[sec:results]{Section 3}, we discuss the simulation results for different accretion rates and column sizes, \lz{compute their time-averaged spatial structures and compare with the \BSmodel}, \lz{discuss the physical origin of the vertical oscillations}, and \lz{discuss a resolution study and the presence of photon bubbles in the simulations.} In \hyperref[sec:discussion]{Section 4}, we discuss the validity of our simulations, compare with previous numerical works, and discuss the observational significance.  \lz{We summarize our conclusions in \hyperref[sec:conclusions]{Section 5}.}

\section{Numerical Method}
\label{sec:numerical_method}

\subsection{Equations}
\label{sec:equations}

Neutron star accretion columns are confined by strong magnetic fields ($10^{9-12}~\mathrm{G}$ or even higher). These strong magnetic fields present challenges for numerical simulations, and all the early work on photon bubble instability adopted equations of motion in which the fluid was confined to move on infinitely rigid field lines, i.e. MHD was not explicitly included \citep{1989ESASP.296...89K, 1996ApJ...457L..85K, 2020PASJ...72...15K}.  Here we include the full MHD dynamics.  One resulting challenge is that the CFL condition on the time step in Newtonian MHD can greatly slow down the numerical simulation because the Newtonian Alfv\'en speed in low density regions (especially outside the column) can easily exceed the speed of light. We therefore employ relativistic MHD because the relativistic Alfv\'en speed is intrinsically limited by the speed of light. 

We apply the numerical framework of radiative \lz{relativistic magnetohydrodynamics} in \textsc{Athena++} \citep{Zhang2021} to simulate a magnetically confined, accreting column near the neutron star surface in Cartesian geometry. The governing equations are summarized below in the sequence of particle number conservation, momentum conservation, energy conservation and radiative transfer
\begin{subequations}
\begin{align}
    &\partial_0(\rho u^0) + \partial_j(\rho u^j) = S_{\mathrm{gr}1}
    \quad, 
	\label{eq:particle_conserv}
	\\
	&\begin{multlined}[t]
	\partial_0(w u^0u^i - b^0b^i)
	\\
	+ \partial_j\left(w u^iu^j + \left(P_g+ \frac{1}{2}b_{\nu}b^{\nu}\right)\delta^{ij} - b^ib^j \right) = S_{\mathrm{gr}2}^i - S_{r2}^i
    \quad, 
	\end{multlined}
	\label{eq:mom_conserv}
	\\
	&\begin{multlined}[t]
	\partial_0\left[w u^0u^0 - \left(P_g+ \frac{1}{2}b_{\nu}b^{\nu}\right) - b^0b^0\right]
	\\
	\mkern155mu + \partial_j(w u^0u^j - b^0b^j) = S_{\mathrm{gr}3} - S_{r3}
    \quad, 
	\end{multlined}
	\label{eq:energy_conserv}
	\\
	&\partial_0I + n^j\partial_j I = \mathcal{L}^{-1}(\bar{S}_r)
    \quad, 
	\label{eq:rad_transfer}
\end{align}
\end{subequations}
where $\rho$ is the fluid frame gas density and $u^{\mu}$ is the fluid four-velocity. Given the fluid three-velocity $v^i$, we can calculate $(u^0, u^i)=\Gamma(1, v^i)$ in special relativity, where $\Gamma=\left(1-v_jv^j\right)^{-1/2}$ is the Lorentz factor. The total fluid frame enthalpy $w$ and the magnetic field four-vector $b^{\mu}$ are defined as follows
\begin{align}
    b^0 &= u_jB^j,\quad b^i = \frac{1}{\Gamma}(B^i + b^0u^i)
    \quad, 
    \\
    w &= \rho +\frac{\gamma}{\gamma-1}P_g +b_{\nu}b^{\nu}
    \quad, 
    \label{eq:total_enthalpy}
\end{align}
where $B^i$ is the magnetic field three-vector and $P_g$ is the fluid frame gas pressure. We adopt an ideal gas with adiabatic index $\gamma=5/3$. The quantities $S_{\rm gr1}$, $S^i_{\rm gr2}$ and $S_{\rm gr3}$ are weak field gravitational source terms (see Appendix B of \citealt{Zhang2021} for details).

In the radiative transfer equation \autoref{eq:rad_transfer}, $I$ is the frequency-integrated, but angle-dependent intensity, and the unit vector $n^i$ is the photon propagation direction. We adopt 40 radiation angles, which are in the plane of the 2D simulation. This is the same angular discretization that we used in \citet{Zhang2021}, and the details are described in \citet{2021ApJS..253...49J}.
Note that the radiative transport source term is initially computed in the fluid frame ($\bar{S}_r$) and then Lorentz transformed back to the lab frame ($S_r$).\footnote{\lz{Throughout this paper, radiation quantities in the fluid rest frame are indicated with a bar.}}
The operator $\mathcal{L}$ represents a Lorentz boost from the lab frame to the fluid frame and $\mathcal{L}^{-1}$ is its inverse. Therefore, $S_r=\mathcal{L}^{-1}(\bar{S}_r)$ and we have source terms in the lab frame exchanging momentum and energy between gas and radiation as follows
\begin{subequations}
\begin{align}
    S_{r2}^i &= \oint S_r n^i d\Omega
    \quad, 
    \\
    S_{r3} &= \oint S_r d\Omega
    \quad, 
\end{align} 
where $d\Omega$ is the infinitesimal solid angle about the photon propagation direction $n^i$. The radiative transport source term in the fluid frame \citep{2021ApJS..253...49J, Zhang2021} is 
\begin{align}
    \begin{split}
        \bar{S}_r = \Gamma(1-v_jn^j)\bigg[ &\rho\kappa_s(\bar{J}-\bar{I})
        \\
        +&\rho\kappa_R\left(\frac{a_r T_g^4}{4\pi}-\bar{I}\right) +\rho(\kappa_P-\kappa_R)\left(\frac{a_r T_g^4}{4\pi}-\bar{J}\right)
        \\
        +&\rho\kappa_s\frac{4(T_g-\bar{T}_r)}{T_e}\bar{J} \bigg]
    \quad, 
    \label{eq:rad_source_term_fluid_frame}
    \end{split}
\end{align}
\end{subequations}
where $\bar{I}=\mathcal{L}(I)$ is the fluid-frame frequency-integrated intensity and $\bar{J} = (4\pi)^{-1}\oint \bar{I} d\bar{\Omega}$ is the fluid-frame zeroth angular moment of intensity. Other quantities are the electron scattering opacity $\kappa_s$, Rosseland mean absorption opacity $\kappa_R$, Planck mean absorption opacity $\kappa_P$, radiation density constant $a_r$, effective temperature of the radiation $\bar{T}_r = (4\pi\bar{J}/a_r)^{1/4}$, and electron rest mass energy expressed as a temperature \lz{$T_e\equiv m_{\rm e}c^2/k_{\rm B}$}. The gas temperature is $T_g=P_gm/(\rho k_{\rm B})$, \lz{where $m$ is the solar abundance mean molecular mass}. In the source term \autoref{eq:rad_source_term_fluid_frame}, the factor \lz{$\Gamma(1-v_jn^j)$} outside the square brackets comes from the frame transformation, the first term in the square brackets refers to elastic scattering, the second and third terms represent absorption and emission processes, and the last term is an approximation for the gas-radiation heat exchange via Compton scattering (\citealt{2003ApJ...596..509B}; \citealt{2009ApJ...691...16H}).

Throughout this paper we assume that the opacity is nonmagnetic isotropic Thomson scattering without polarization dependence. We also adopt a fully grey treatment of radiation transfer with an assumed blackbody spectrum, neglecting nonzero photon chemical potential effects.  Magnetic scattering opacities in particular are likely to be important in the dynamics of actual accretion columns with stronger magnetic fields than we have employed here, and we intend to explore this in future papers.  The simulations we present here will provide a physics baseline for these future investigations.

\subsection{Simulation Domain}
\label{sec:simulation_setup}

\begin{figure}
    \centering
	\includegraphics[width=\columnwidth]{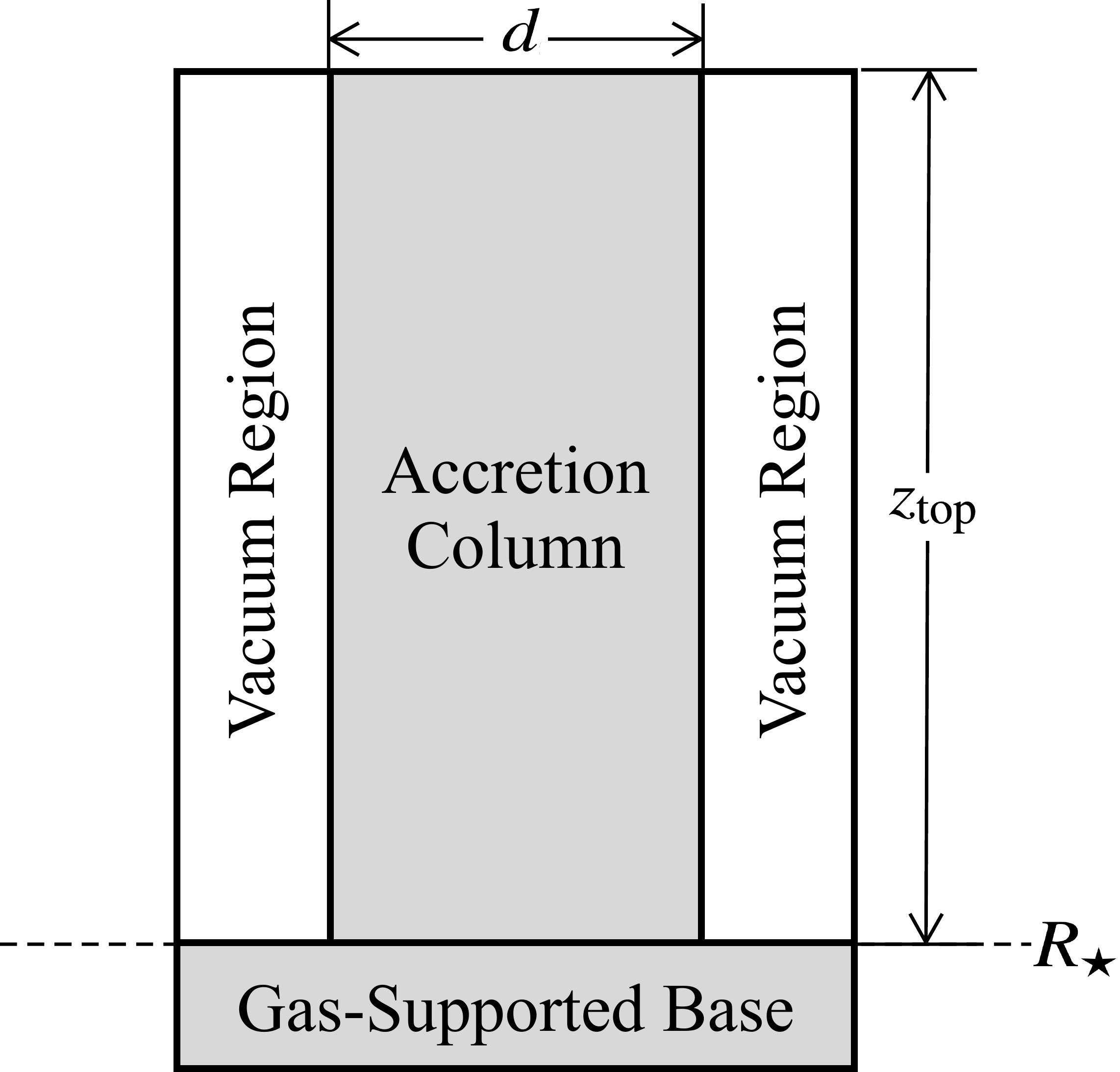}
    \caption{Partitioning of the simulation domain into different regions with distinct numerical treatments.  The region labelled accretion column is the active, physical region where material accretes through the top boundary.  On either side of this column is a \lz{vacuum} region through which no material enters from the top, and in which an effective boundary condition is maintained by setting the density \lz{and gas pressure to their floor values.}  Underneath all these regions is a gas-pressure supported base with an artificially reduced gravitational acceleration in order to provide adequate spatial resolution, and which provides an effective base boundary condition.  (See \autoref{sec:effective_bc} for details of all these effective boundary conditions.)}
    \label{fig:column_partition}
\end{figure}

As we discovered earlier in \citet{Zhang2021}, the nonlinear behavior of the photon bubble instability is resolution-dependent and dominated by the shortest resolved wavelengths above the radiation viscous scale. Therefore, we prefer to adopt a Cartesian column initialized with a uniform vertical magnetic field so that the simulation can be configured on a uniform grid, which avoids the extra complexity introduced by the geometric dilution in any other kind of geometry. Such geometric dilution (e.g. in spherical or dipolar geometry) can naturally lead to non-uniform numerical behavior of the photon bubble instability, in particular by not resolving the short-wavelength faster growing modes at higher altitudes. However, this Cartesian geometry limits what parameter regimes we can simulate: the accretion rates in our simulations should be small enough that the height of the shock front is small compared to the stellar radius.

Our simulations are confined to the 2D $x$-$z$ plane, and gas only moves in this plane. However, the radiation intensity propagation directions $n^j$ are fully 3D.  There are no spatial gradients in the $y$-direction perpendicular to the $x$-$z$ plane, implying that there is also no $y$-component of radiation flux.  Our 2D simulations can be interpreted as a vertical slice of a narrow wall-shaped column, e.g. as shown in the Fig.I of \citet{1976MNRAS.175..395B}. 

The simulation domain is partitioned as illustrated in \autoref{fig:column_partition}. 
We define $z_{\mathrm{top}}$ as the height at the top of the simulation domain and $d$ as the width at the bottom of the accretion column.  (Again, $d$ should be interpreted as the width of a narrow-walled column which is much shorter than the extent of the column in the $y$-direction perpendicular to the $x$-$z$ plane.  We are neglecting all spatial gradients in that $y$-direction in our 2D simulations.)
Note that we add a gas-supported layer at the base and vacuum regions at the two sides as effective boundaries of the accreting column for numerical purposes, which will be discussed below in \autoref{sec:effective_bc}.

We parameterize the accretion rate based on the effective luminosity $L_{\mathrm{eff}}$ and Eddington flux ratio $\epsilon$ respectively, where the effective luminosity is the area-weighted Eddington luminosity and the Eddington flux ratio is the ratio of the accretion luminosity $L_{\mathrm{acc}}$ to the effective Eddington luminosity. Given the Eddington luminosity $L_{\mathrm{Edd}}=4\pi GM_{\star}c/\kappa_{s}$ of a neutron star with mass $M_{\star}=1.4M_{\sun}$ and radius $R_{\star}=10^6~\mathrm{cm}$, we have 
\begin{align}
    L_{\mathrm{eff}} &= \frac{A}{4\pi R_{\star}^2}L_{\mathrm{Edd}}
    \quad,
    \\
    L_{\mathrm{acc}} &= \frac{GM_{\star}\rho_{\mathrm{acc}}Av_{\mathrm{ff}}}{R_{\star}} = \epsilon L_{\mathrm{eff}}
    \label{eq:rho_acc}
    \quad,
\end{align}
where $A$ is the \lz{horizontal} area of the accretion column on the neutron star surface and $\rho_{\mathrm{acc}}$ is the accreting gas density at the free-fall speed \lz{$v_{\mathrm{ff}}=\sqrt{2GM_{\star}/(R_{\star}+z_{\rm top})}$ at the top of the simulation domain.}  Therefore, we can tune the accretion rate in terms of $\rho_{\mathrm{acc}}$ in the simulation by simply adjusting $\epsilon$.

\subsection{Initial Conditions}
\label{sec:accreting_column}

In standard models of neutron star accretion columns, which are based on the seminal work of \citet{1976MNRAS.175..395B}, the magnetic fields are strong enough to constrain the inflow to move along the field lines near the stellar surface. At high accretion rates, the incoming material free falls onto the neutron star and forms a shock above the stellar surface, where the region above the shock front is called the free-fall zone and the region below is called the sinking zone. At the bottom of the free-fall zone, the inflow is halted by the shock front which converts nearly all its kinetic energy into radiation energy. The sinking zone is radiation-dominated regardless of the magnetic pressure, where the gas is in approximate hydrostatic equilibrium between upward \lz{radiation force} and downward gravity, with a slow subsonic sinking flow toward the neutron star surface.  We follow this configuration to initialize our 2D numerical simulation.

In the beginning of the simulation, we assume that the vertical magnetic fields are strong and uniform across the whole domain, where $B_{x0}=0$ and $B_{z0}=8\times10^{10}~\mathrm{G}$, where the subscript 0 refers to the initial condition. While this is somewhat below the $10^{12}$~G, or more, magnetic fields that are typical of high mass X-ray binary pulsars, it is large enough to rigorously confine the material in the accretion column.  Higher fields will not affect the dynamics, and create numerical problems as they have energy densities which are too large compared to the gas pressure (see \autoref{sec:validity_param} for details).  It is true that higher fields can reduce the scattering opacity, but again, we are assuming constant Thomson opacity here (and the Planck and Rosseland mean absorption opacities have negligible effects), and intend to explore the magnetic opacity effects in a future paper. 

Although we use radiative \lz{relativistic MHD} in \textsc{Athena++} \citep{2011ApJS..193....6B, 2021ApJS..253...49J, Zhang2021} for the numerical simulation, only the free-fall region of the column is mildly relativistic ($\Gamma\simeq1.31$) and the initial magnetic field is strong enough to \lz{enforce purely vertical motion of the gas}. Therefore, we formulate the initial condition using Newtonian physics and assuming 1D gas motion along the magnetic fields for simplicity.  The time-independent conservation laws for the initial condition can then be written as follows:

\begin{subequations}
\begin{align}
    & \frac{\partial}{\partial z}\left(\rho_0 v_{z0}\right) = 0
    \quad,
    \label{eq:time_indep_mass_conserv}
    \\
    & \rho_0 v_{z0}\frac{\partial v_{z0}}{\partial z} = -\frac{\partial P_{g0}}{\partial z} - \rho_0\left(g(z)-\frac{\kappa}{c}\bar{F}_{z0}\right)
    \quad,
    \label{eq:time_indep_momentum_conserv}
    \\
    \begin{split}
    & v_{z0}\frac{\partial}{\partial z}\left(\frac{1}{\gamma-1}P_{g0}+\bar{E}_{r0}\right) + \left(\frac{\gamma}{\gamma-1}P_{g0}+\bar{E}_{r0}+\bar{P}_{r0}\right)\frac{\partial v_{z0}}{\partial z}
    \\
    &\mkern250mu = -2\frac{\partial \bar{F}_{x0}}{\partial x} - \frac{\partial \bar{F}_{z0}}{\partial z}
    \quad, 
    \label{eq:time_indep_energy_conserv}
    \end{split}
\end{align}
\label{eq:time_indep_consev}
\end{subequations} \\
where the flux mean opacity $\kappa=\kappa_s+\kappa_R$ is the effective opacity for gas-radiation momentum coupling and $g(z)=GM_{\star}(R_{\star}+z)^{-2}$ is the gravitational acceleration.  The factor of 2 before the $\partial \bar{F}_{x0}/\partial x$ term in \eqnref{eq:time_indep_energy_conserv} comes from the radiative cooling on both sides of the column.  We then follow the approximations in \citet{1976MNRAS.175..395B} to simplify equations~\autoref{eq:time_indep_consev} and compute the 1D solution along the vertical direction ($\hat{z}$) to initialize the sinking zone. We apply the following approximations for the gas and radiation field: 
\begin{enumerate}
  \item Local thermal equilibrium (LTE) between the gas temperature and radiation effective temperature $T_{g0}=\bar{T}_{r0}\equiv T_0$. 
  \item Standard Eddington closure scheme between the zeroth and second angular moments of the radiation field 
  \begin{equation}
      \mkern175mu \bar{P}_{r0}=\frac{1}{3}\bar{E}_{r0}
      \quad.
  \end{equation}
  \item Radiative transport in the diffusion approximation 
  \begin{equation}
      \mkern150mu \bar{F}_{x0} = -\frac{c}{\rho_0\kappa}\frac{\partial \bar{P}_{r0}}{\partial x}
      \quad.
  \end{equation}
  \item We approximate horizontal gradients with a simple one zone finite difference between the center and each side of the column, assuming $\bar{F}_{x0}=0$ at the center of the column and $\bar{P}_{r0}=0$ at the sides.  E.g. for the right half:
  \begin{equation}
      \mkern75mu
      \dfrac{\partial \bar{F}_{x0}}{\partial x} \simeq \dfrac{2\bar{F}_{x0}}{d}
      \quad\textrm{and}\quad
      \dfrac{\partial \bar{P}_{r0}}{\partial x} \simeq -\dfrac{2\bar{P}_{r0}}{d}
      \quad.
  \end{equation}
\end{enumerate}
Recall that $d$ is the column width.  Therefore, the time-independent conservation laws \autoref{eq:time_indep_consev} can be further reduced and rearranged for the convenience of numerical integration as follows 
\begin{subequations}
\begin{align}
    & \frac{\partial T_0}{\partial z} = -\frac{3\rho_0\kappa}{4a_r T_0^3}\frac{\bar{F}_{z0}}{c}
    \quad,
    \label{eq:sinking_profile_temp}
    \\
    & \frac{\partial\rho_0}{\partial z} = \dfrac{-\rho_0\left( g(z)-\dfrac{\kappa}{c}\bar{F}_{z0}+\dfrac{k_{\rm B}}{m}\dfrac{\partial T_0}{\partial z} \right)}{(k_{\rm B}T_0/m)-v_{z0}^2}
    \quad,
    \label{eq:sinking_profile_density}
    \\
    & \frac{\partial v_{z0}}{\partial z} = \frac{s_0}{\rho_0^2} \frac{\partial\rho_0}{\partial z}
    \quad,
    \label{eq:sinking_profile_velocity}
    \\
    \begin{split}
        & \frac{\partial \bar{F}_{z0}}{\partial z} = -\frac{4c a_r T_0^4}{3\rho_0\kappa d^2} - v_{z0} \left(\frac{3}{2}\dfrac{k_{\rm B}}{m}\rho_0+4a_rT_0^3\right) \frac{\partial T_0}{\partial z}
        \\
        &\mkern75mu - \frac{3}{2}\dfrac{k_{\rm B}}{m}v_{z0}T_0\frac{\partial\rho_0}{\partial z} - \left(\frac{5}{2}\dfrac{k_{\rm B}}{m}\rho_0 T_0+\frac{4}{3}a_r T_0^4\right)\frac{\partial v_{z0}}{\partial z}
        \quad,
    \end{split}
    \label{eq:sinking_profile_radflux}
\end{align} \label{eq:ic_numerical}
\end{subequations} \\
where the mass accretion rate per unit area $s_0=-\rho_0 v_{z0}$ is a constant.
After the simulation domain and accretion parameters are determined, we can obtain the gas density in the free-fall region via \eqnref{eq:rho_acc} and then the free-fall speed $v_{\rm ff}$.  Given that the gas is cold ($5\times10^6~\mathrm{K}$, following \citealt{1989ESASP.296...89K}) and the radiation flux is negligible at the top, we can then continue the free-fall solution down to a specified height $z_{\mathrm{int}}$ which is high enough to be above the accretion shock and sinking region.  We then integrate equations~\autoref{eq:ic_numerical} downward until the vertical speed at the bottom \lz{decreases to a value} $\sim 10^{-3}c$. 

\subsection{Boundary Conditions}
\label{sec:effective_bc}

As illustrated in \autoref{fig:column_partition}, besides the accretion column region, we also add a gas-supported layer at the base and two vacuum regions
on both sides. These regions are artificial but necessary to serve as effective boundary conditions (hereafter soft boundaries) for better numerical performance compared with the direct implementation on the boundaries of the simulation domain (hereafter hard boundaries). 

At the lower boundary of the simulation domain, we zero the horizontal components of the magnetic field while maintaining constant vertical component.  We also apply reflective boundary conditions for both the gas and the radiation.  Such a hard reflective boundary condition for the gas would inevitably lead to gas leakage out of the accreting column at an unphysically high rate since it simply copies the horizontal components of the velocity in the ghost zones, which lack magnetic confinement because we enforce purely vertical magnetic field there. To alleviate this problem, we introduce a thick gas-supported layer as an effective boundary for the sinking gas in the accretion column. 

This gas-supported base mocks up the presence of the neutron star surface within the simulation domain.  We initialize it to be cold ($T_{g0}=\bar{T}_{r0}=T_{\mathrm{b}}=5\times10^6~\mathrm{K}$) and optically thick so that radiation pressure is negligible.  It is fully supported against gravity by gas pressure but still constrained by the magnetic field. A realistic gas pressure scale height in the gravitational field of a neutron star is very small ($\sim3.7~\mathrm{cm}$), and would require unnecessarily high resolution merely for an effective boundary.  We therefore artificially reduce the gravitational acceleration by a factor of 100 in the gas-supported layer to increase the scale height so that we do not waste too many grid zones.  We initialize the gas-supported base in hydrostatic equilibrium and determine the gas density via $\rho_{\mathrm{b}}\propto\mathrm{exp}(-z/h_{\mathrm{b}})$, where $h_{\mathrm{b}}=RT_{\mathrm{b}}/g_{\mathrm{b}}(\sim370~\mathrm{cm})$ is the scale height and $g_{\mathrm{b}}=GM_{\star}/(100R_{\star}^2)$ is the reduced gravitational acceleration at the neutron star surface. 

The side boundary conditions also require careful treatment. In contrast to the static neutron star atmospheres that we simulated in \citet{Zhang2021}, where we used horizontally periodic boundary conditions, here we must use side boundary conditions that permit the formation of an accretion column that 1. allows for escape of photons from the side and 2. is horizontally confined by the magnetic field against the resulting sideways radiation pressure forces.  We allow for the escape of photons by zeroing the intensity along all inward pointing propagation angles at the side boundaries.  Barring any instabilities, the magnetic field we use is strong enough (\lz{with energy density} at least 100 times larger than the total thermal pressure) that horizontal confinement is easily achieved.  However, we have conducted numerical experiments with an outflow boundary condition for the gas on the side boundaries, and these always lead to escape of gas that drags the field along with it.  This is simply because we did not impose an inward force at the boundary to compensate for the outward thermal pressure forces.  Rather than do that, we instead impose a reflecting boundary condition at the side boundaries.  As these boundaries are displaced from the actual accretion column, this still allows for dynamic confinement of the column by the magnetic field.  In order to allow photons to horizontally escape from the actual accretion column, we also reset the density and gas temperature in the side regions to the floor values at each time step to mock up real vacuum conditions.  These effective side boundary conditions have three major advantages: 
\begin{enumerate}
    \item We avoid the artificial accretion associated with an infalling density floor. 
    \item We guarantee that the vacuum regions are optically thin so that the radiation can freely leave the sides of the accretion column. 
    \item We prevent artificial gas ejection from the gas-supported base (where we have imposed artificial low gravity) in the side regions caused by irradiation heating from the accretion column.
\end{enumerate}

In addition to our treatment in these effective side and bottom boundary regions, we still have to specify the boundary conditions at the actual edges of our simulation domain.  We adopt an upper boundary condition for the magnetic field that zeroes out the horizontal component ($B_x=0$) and maintains constant vertical magnitude ($B_z=8\times10^{10}~\rm{G}$) in the ghost zones. The simulations are fed with cold ($5\times10^6~\mathrm{K}$) and optically thick material with density $\rho_{\mathrm{acc}}$ that varies between simulations through the top boundary.  This material free falls at speed $v_{\mathrm{ff}}$ into the simulation domain through the upper boundary of the free-fall zone.  At both sides of the simulation domain, we adopt vacuum boundary conditions for the radiation (see Section 3.2 in \citealt{Zhang2021} for details) and reflective boundary conditions for both the gas and the magnetic field.  At the \lz{very bottom, as noted previously,} we use reflective boundary conditions for the radiation, where the intensity is specularly reflected from the boundary. The gas boundary condition is also reflective but the magnetic field is forced to be constant just like the upper boundary condition.

\subsection{Issues with Variable Inversion} 
In addition, although we update the primitive variable inversion algorithm as described in \citet{Zhang2021}, it still fails in most low-density regions where the gas density is too low and the orders of magnitude difference between magnetic pressure and gas pressure causes difficulty in determining the gas pressure from the conservative total energy density.  Because gas pressure is dynamically unimportant, this would not normally matter.  However, in low density regions, the noisily determined gas pressure leads to noise in the gas temperature and occasional unphysically high gas temperature regions.  The radiation field would then be heated by this numerical noise because the gas temperature is used to compute the gas-radiation energy exchange source term (which is dominated by Compton scattering). In order to control such artificial heating of the radiation, we apply a density threshold $\rho_{\mathrm{comp}}$ which is selected to be the same as $\rho_{\mathrm{acc}}$ and only allow Compton scattering when the local density is greater than $\rho_{\mathrm{comp}}$.  However, this ad hoc fix works for the simulations presented here, but alternative approaches will need to be developed for strongly magnetized gas for simulations in more extreme regimes (e.g. global simulations where the accreting gas can be wind-captured, fed from a disk or the magnetar regime where the magnetic field strength can reach $10^{13-15}~\mathrm{G}$). 

\subsection{Simulation Parameters}
\label{sec:three_sim_versions}

\begin{table*}
	\centering
	\begin{tabular}{cccccccccc}
		\hline
		Version & Name & Mesh & $d$ & $z_{\mathrm{int}}$ & $z_{\mathrm{top}}$ & \lz{Resolution} & $\epsilon$ & $\rho_{\mathrm{acc}}$\\
% 		&  &  & ($R_{\star}$)  & ($R_{\star}$) & ($R_{\star}$) & ($\mathrm{cm^2/cell}$) & & ($10^{-4}~\mathrm{g/cm^3}$)\\
		&  &  & ($R_{\star}$)  & ($R_{\star}$) & ($R_{\star}$) & ($10^{-3}\mathrm{cells/cm}$) & & ($10^{-4}~\mathrm{g/cm^3}$)\\
		\hline
% 		0 & HR-Narrow-100 & $192\times1024$ & 0.015 & 0.100 & 0.150 & $156\times154$ & 100 & 4.60\\
% 		1 & HR-Narrow-150 & $200\times2400$ & 0.015 & 0.125 & 0.350 & $146\times148$ & 150 & 6.90\\
% 		2 & HR-Wide-25 & $700\times2048$ & 0.060 & 0.340 & 0.350 & $171\times173$ & 25 & 1.15\\
% 		3 & MR-Wide-25 & $350\times1024$ & 0.060 & 0.340 & 0.350 & $342\times346$ & 25 & 1.15\\
% 		4 & LR-Wide-25 & $176\times512$  & 0.060 & 0.340 & 0.350 & $686\times692$ & 25 & 1.15\\
		0 & HR-Narrow-100 & $192\times1024$ & 0.015 & 0.100 & 0.150 & 6.5 & 100 & 4.60\\
		1 & HR-Narrow-150 & $200\times2400$ & 0.015 & 0.125 & 0.350 & 6.8 & 150 & 6.90\\
		2 & HR-Wide-25 & $700\times2048$ & 0.060 & 0.340 & 0.350 & 5.8 & 25 & 1.15\\
		3 & MR-Wide-25 & $350\times1024$ & 0.060 & 0.340 & 0.350 & 2.9 & 25 & 1.15\\
		4 & LR-Wide-25 & $176\times512$  & 0.060 & 0.340 & 0.350 & 1.5 & 25 & 1.15\\
		\hline
	\end{tabular}
	\caption{Initialization parameters of our five accretion column simulations.}
	\label{tab:sim_param}
\end{table*}

\begin{figure*}
    \centering
	\includegraphics[width=\textwidth]{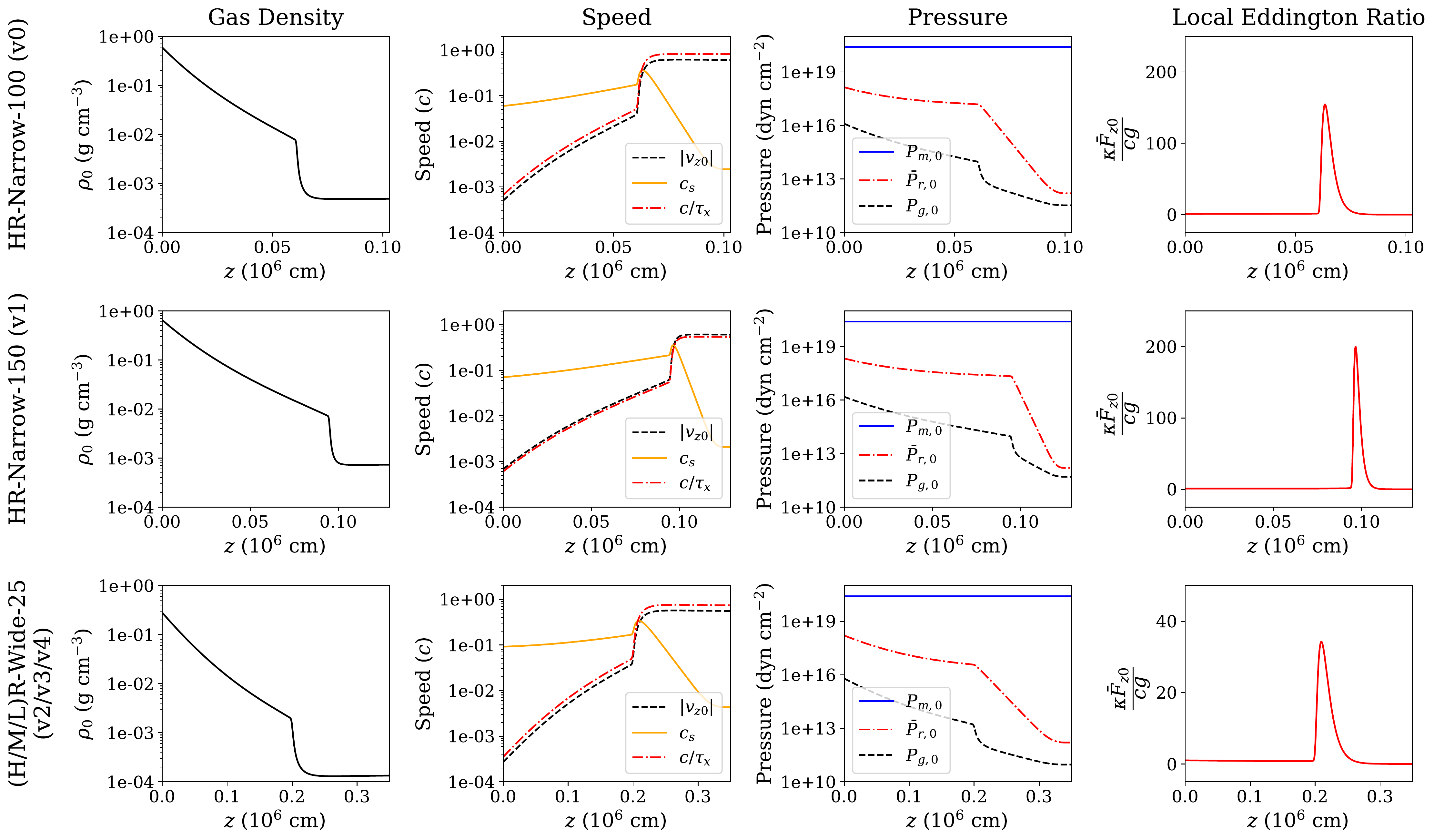}
    \caption{\lz{Initial condition in our simulations, computed using the \BSmodel.} The top and middle row of panels correspond to simulations v0 and v1, while the bottom row corresponds to simulation v2 as well as simulations v3 and v4 used in our resolution study.  From left to right, the panel columns show initial fluid frame density, various speeds (the vertical fluid speed, the sound speed, and the horizontal photon diffusion speed), various pressures (magnetic, radiation, and gas), and the vertical \lz{fluid}-frame radiation flux (expressed as a local Eddington ratio).}
    \label{fig:ini_condition}
\end{figure*}

We have run five accretion column simulations with parameters listed in \autoref{tab:sim_param}. We set up the first three versions (v0, v1, v2) by varying the column width and accretion rate to study how the column dynamics depends on these parameters. We take the first simulation (v0) as our fiducial simulation. We increase the accretion rate by a factor of 1.5 while maintaining the same column width in the second simulation (v1). The third simulation (v2) has the column 4 times wider while maintaining the same mass accretion rate and thus $\epsilon$ is 4 times smaller. These first three simulations all have similar resolutions in both the horizontal and vertical directions. We set up the last two simulations (v3, v4) for a resolution study by decreasing the resolution by factors of 2 and 4, respectively, for the parameters of simulation v2, where the column exhibits a \lz{continually} multi-peaked structure in the horizontal direction. Each simulation runs for $7100t_{\mathrm{sim}}$, where $t_{\mathrm{sim}}=2.8\times10^{-7}~\mathrm{s}$ is the simulation time unit.

The initial vertical profiles of various quantities in the accretion column are shown in \autoref{fig:ini_condition}. The sharp changes in gas density and velocity clearly indicate the location of the shock structure at the top of the sinking zone.  As illustrated in the pressure plots, the column is radiation pressure dominated with strong magnetic confinement. Note that $P_m$ is the magnetic pressure and $c_s$ is the adiabatic sound speed in the radiation and gas mixture, \lz{$\tau_x=\rho_0\kappa d/2$} is the horizontal optical depth of the column \lz{(from the center to the side)} and thus $c/\tau_x$ represents the radiative diffusion speed inside the column in the horizontal direction. 

\section{Results}
\label{sec:results}
In the following subsections, we present and analyze the simulation results. In \autoref{sec:dynamics_column_dynamics}, we summarize the dynamical behavior of the simulations.  \lz{In \autoref{sec:profile_time_ave}, we} compare the time-averaged structures with the classical accretion column model. \lz{In \autoref{sec:interpretation_oscillation} we present a physical interpretation of the vertical oscillations.  In \autoref{sec:resolution_dependence}, we present evidence that photon bubbles are present in at least our highest resolution wide column simulation.}   Animations for all of the simulations are available online\footnote{\url{https://youtube.com/playlist?list=PLbQOoEY0CFpX935unYJWB3gfwhDwcKJ6c}}.

\subsection{Dynamics in Column Accretion Simulations}
\label{sec:dynamics_column_dynamics}
In this section, we use three high-resolution simulations (v0, v1, v2) to illustrate three \lz{different} dynamical behaviors of the accretion column.  We first describe the fiducial simulation (v0) in detail and then discuss the changes that occur at higher accretion rate (v1) and wider size (v2). 

\begin{figure}
    \centering
	\includegraphics[width=\columnwidth]{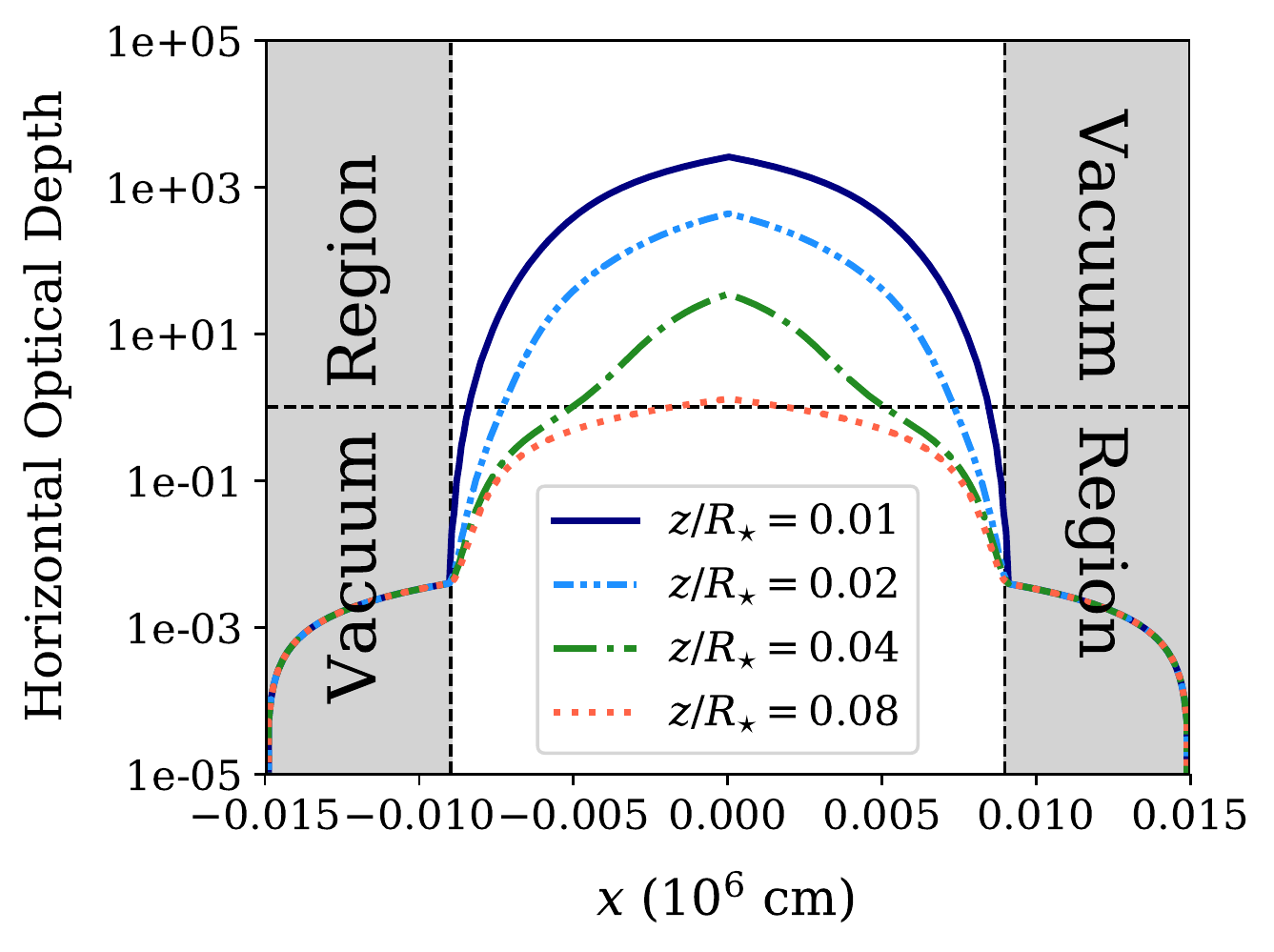}
    \caption{Horizontal optical depths at different altitudes based on the time-averaged profiles of the fiducial simulation v0. The optical depth is calculated from the nearest side of the simulation domain. \lz{The horizontal dashed line indicates optical depth unity, and demonstrates why the accretion column must adopt a round shape.}}
    \label{fig:taux_time_avg_v0}
\end{figure}

\begin{figure*}
    \centering
	\includegraphics[width=\textwidth]{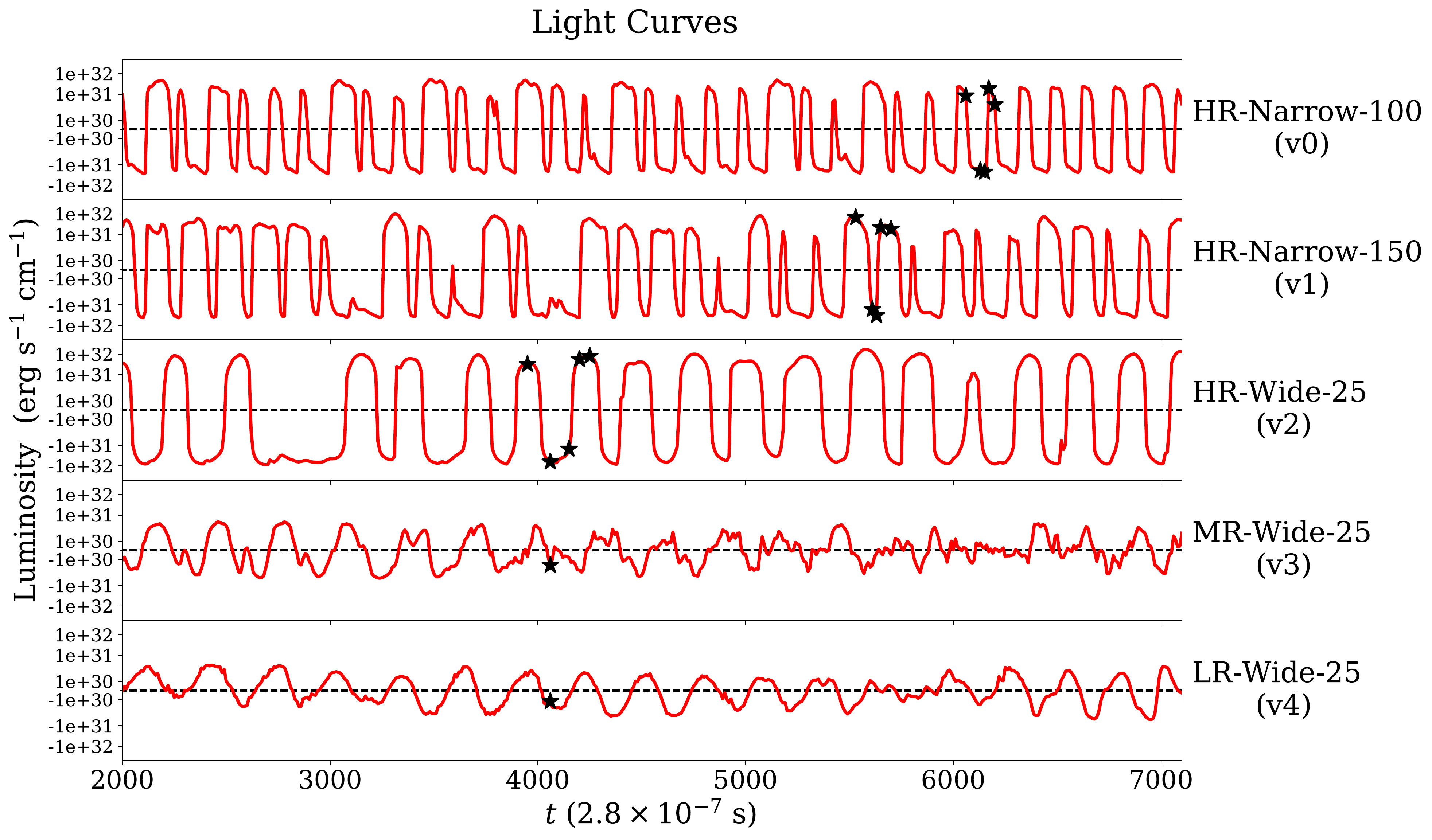}
    \caption{Variations in \lz{vertically integrated sideways} luminosity about the time-average over the time interval depicted here (\lz{after relaxation from the initial condition}) for our five simulations.  The vertical flux leaving the domain is not included in this plot, but it is negligible. Black stars correspond to the snapshots shown in \autoref{fig:density_period_v0}, \autoref{fig:density_period_v1}, \autoref{fig:density_period_v2}, and \autoref{fig:horizontal_incoherence}.  \lz{Oscillations that have a flat-topped profile in the light curves of v0, v1, and v2, are associated with the formation of pre-shocks in the accretion flow.  Such pre-shocks are absent during spike-shaped oscillations.}}
    \label{fig:light_curve}
\end{figure*}

\lz{As can be seen in the animations,} each simulation first relaxes from the \lz{1D} initial condition \lz{which has no horizontal structure.  Radiation quickly escapes from the sides due to the low horizontal optical depth in those regions.}  Hence the height of the shock front drops below the starting position \lz{along the sides} due to less vertical radiation \lz{pressure} support.
This naturally leads to a \lz{curved, mound-shaped} shock front enclosing the sinking region.  \lz{This is illustrated} in \autoref{fig:taux_time_avg_v0}, \lz{where we use} the time-averaged profiles of the fiducial simulation (v0) \lz{to compute} the horizontal optical depth to the nearest side of the simulation domain at different altitudes.

Once the relaxation of the initial condition finishes, the system gradually enters into a state of quasi-periodic \lz{vertical} oscillation, which persists through the end of \lz{all five of our simulations}. \lz{As illustrated in \autoref{fig:light_curve} and in the online animations, these oscillations result in luminosity variations. Only the horizontal flux is depicted in this figure, as the flux escaping vertically is smaller by factors of \lz{at most}
$1.3\times 10^{-3}$, and is therefore negligible.  We explain the physical origin of these oscillations in \autoref{sec:interpretation_oscillation} below.}

During these oscillatory epochs, the system continues to be heated by the dissipation of the kinetic energy of the incoming accretion flow at the shock front.  That dissipated energy is rapidly \lz{transferred to the radiation field} via the Compton process, which is then balanced by radiation diffusion, advection, and radiative cooling at the two sides of the column. The column below the shock is supported against gravity by the radiation pressure \lz{gradient}, with \lz{the radiation diffusion time being much longer than the sound crossing time}.  With the exception of the existence of oscillations, all these properties are qualitatively consistent with the high-accretion rate \lz{static accretion column described by} \citet{1976MNRAS.175..395B}.  \lz{As we discuss in detail in \autoref{sec:interpretation_oscillation} below, however, such a static structure has a much longer vertical heat transport time compared to the sideways cooling time in the upper regions, and this must produce vertical oscillations in the column.}

For each high-resolution simulation, we select 5 snapshots to illustrate the quasi-periodic behavior and examine their 2D profiles when the sinking region is most elongated or compressed to understand the column dynamics. 

\subsubsection{Oscillatory behavior in narrow accretion columns}
\label{sec:osc_narrow_column}

\begin{figure*}
    \centering
	\includegraphics[width=\textwidth]{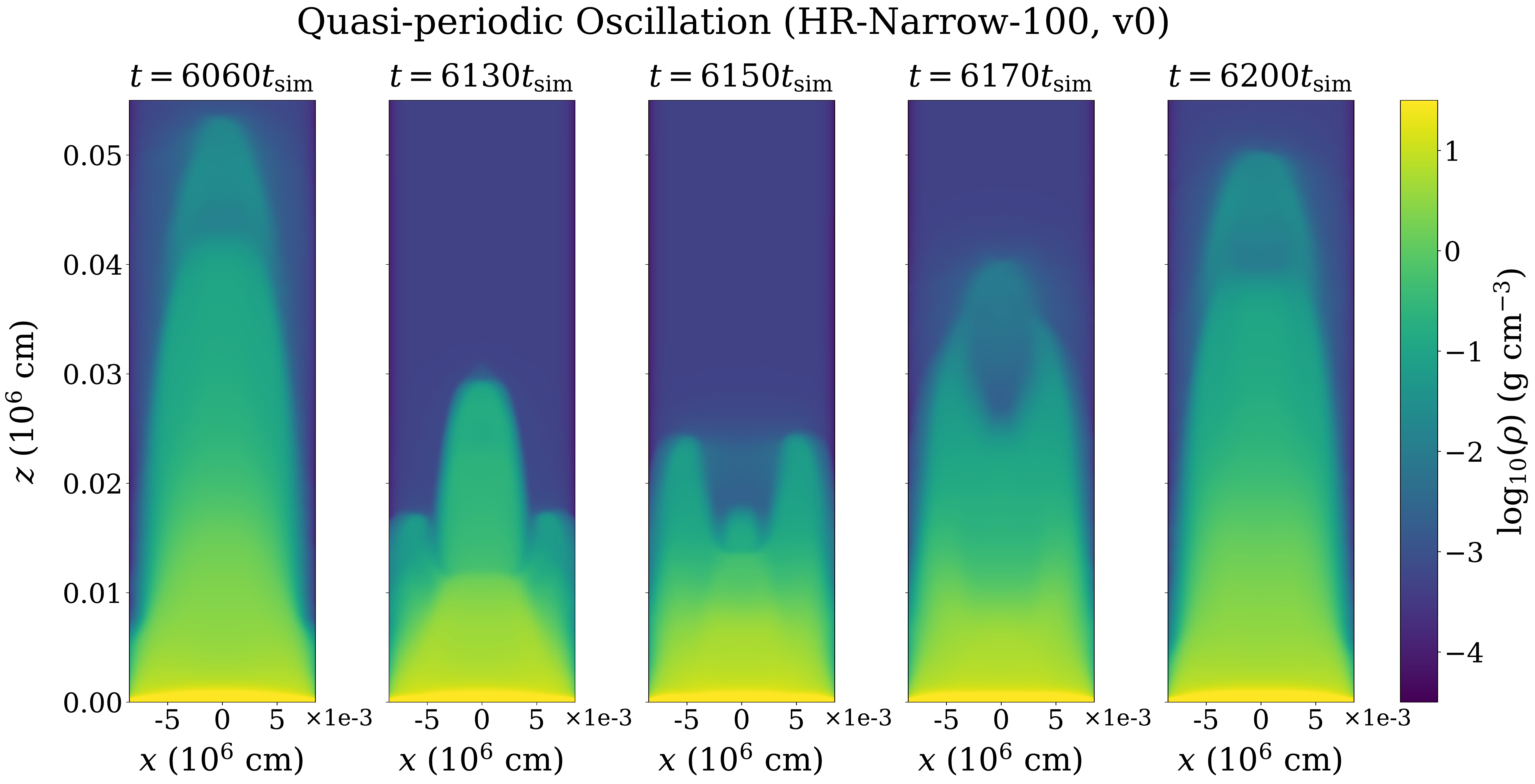}
    \caption{Selected 2D density snapshots \lz{over one full oscillation period $\simeq 140t_{\rm sim}=3.9\times10^{-5}$~s} in the fiducial simulation (v0).  \lz{2D profiles of other fluid quantities at the epochs of maximum (6060$t_{\rm sim}$) and minimum (6150$t_{\rm sim}$) vertical extent are shown in \autoref{fig:profile_snapshot_v0_crest} and \autoref{fig:profile_snapshot_v0_trough}, respectively.}
    }
    \label{fig:density_period_v0}
\end{figure*}

\begin{figure*}
    \centering
	\includegraphics[width=0.90\textwidth]{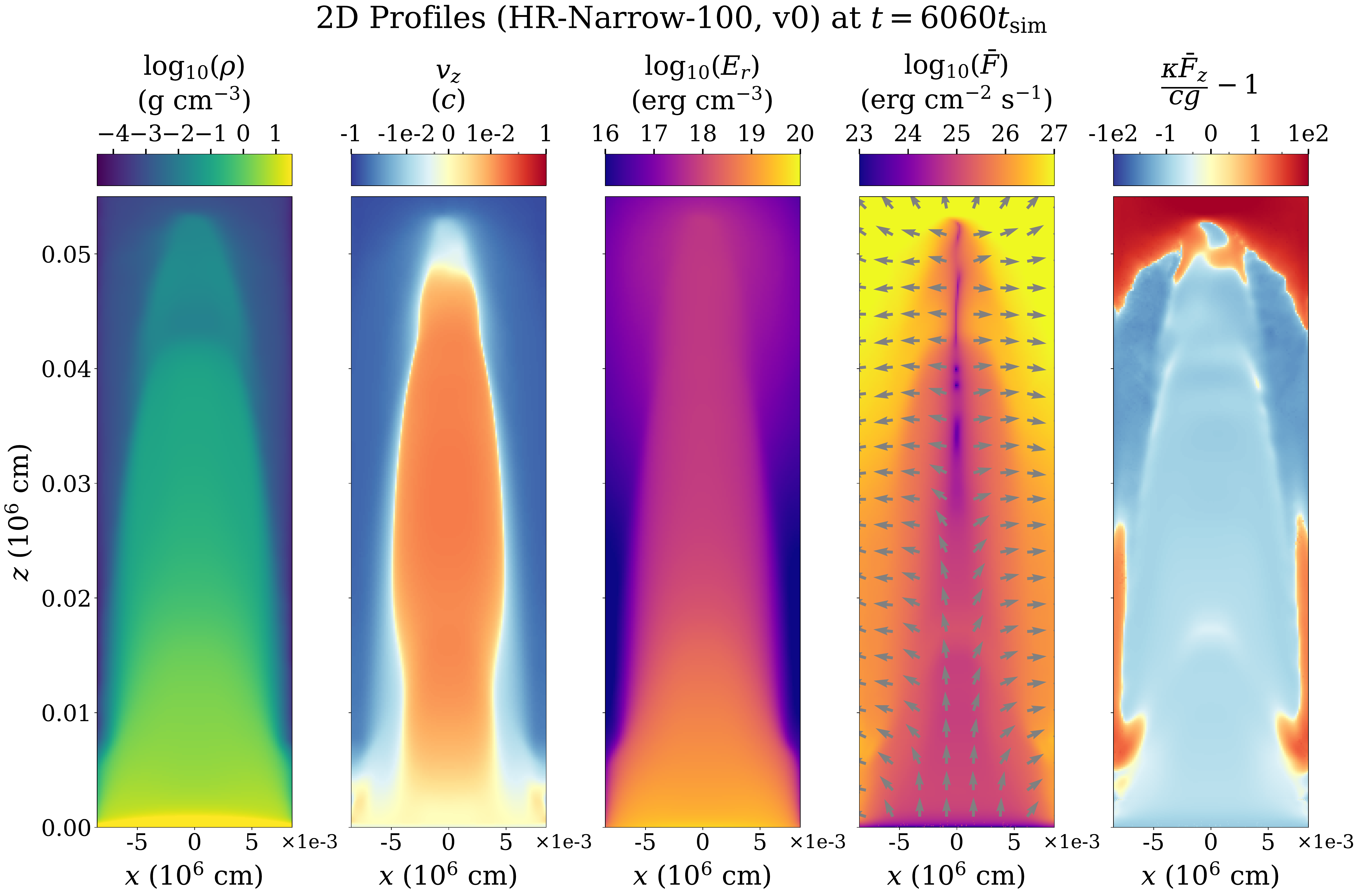}
    \caption{2D profiles at $t=6060t_{\mathrm{sim}}$ in the fiducial simulation (v0). Note that for the vector variables, the arrows represent the direction and the color bar indicates the magnitude.}
    \label{fig:profile_snapshot_v0_crest}
\end{figure*}

\begin{figure*}
    \centering
	\includegraphics[width=0.90\textwidth]{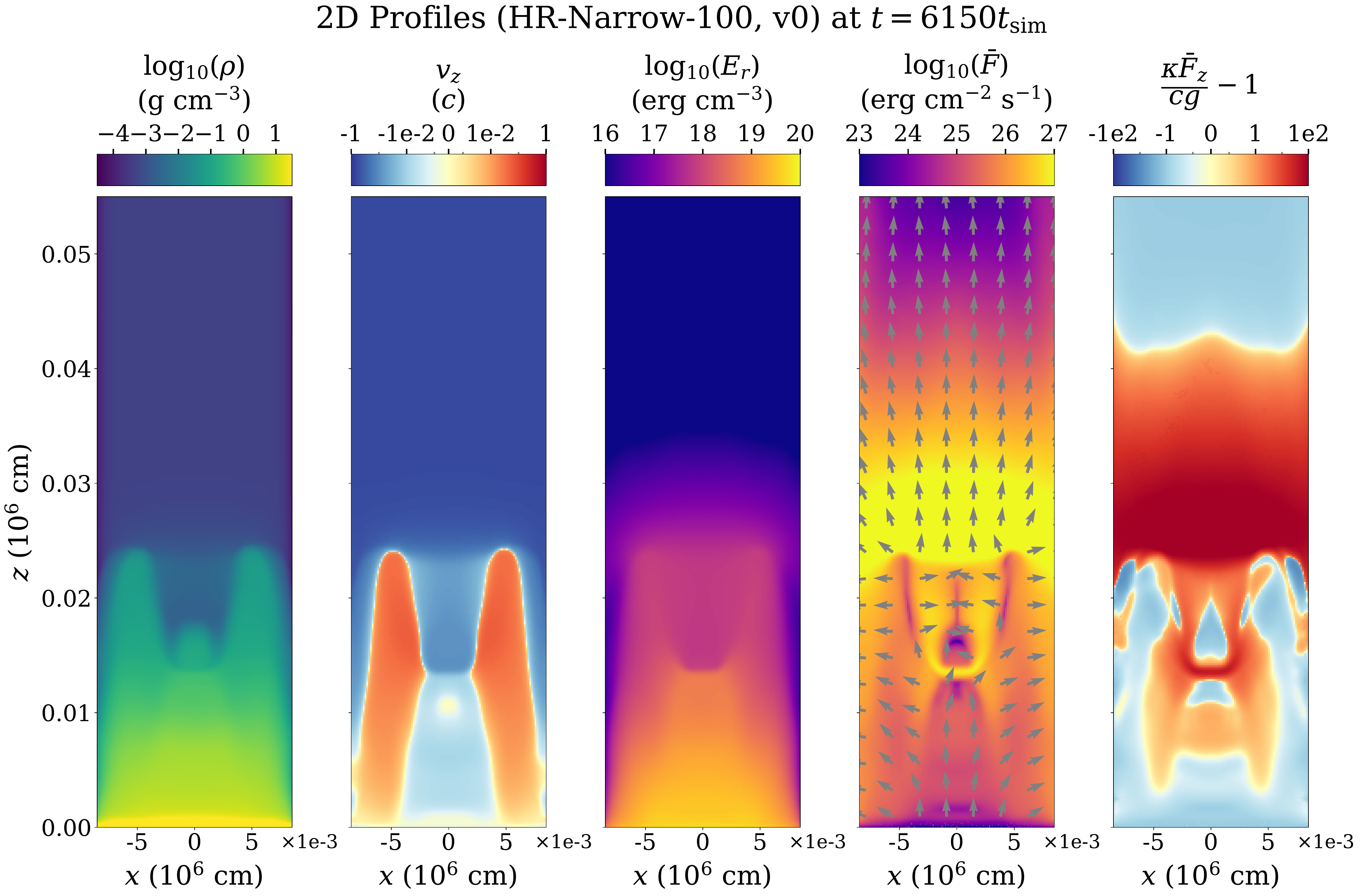}
    \caption{2D profiles at $t=6150t_{\mathrm{sim}}$ in the fiducial simulation (v0). Note that for the vector variables, the arrows represent the direction and the color bar indicates the magnitude.}
    \label{fig:profile_snapshot_v0_trough}
\end{figure*}

The spatial distribution of fluid-frame density over one period of the quasi-periodic oscillation (indicated by black stars in \autoref{fig:light_curve}) in the fiducial simulation (v0) is shown in \autoref{fig:density_period_v0}. The \lz{nonlinear oscillations of the} center and two sides of the column \lz{are sometimes moving in opposite directions during the course of an oscillation period.  This} period ($\sim5\times10^{-5}~\mathrm{s}$) is extremely short and the oscillation amplitude decreases from the center towards the sides. During the oscillation, most of the gas is enclosed by the shock in the sinking region, which is optically thick and radiation dominated. Accretion power continues to heat the shock front.  Bounded by the shock front, a portion of newly generated photons directly escapes out of the column into the vacuum regions and the rest are redistributed through the sinking region, largely through advection by the oscillating column structure.  \lz{Because of the difference in vertical oscillation amplitude between the center and the sides, the column transitions between a symmetric single mound at maximum vertical extent to a triple-peaked structure at minimum vertical extent.}  We shall see that the \lz{high resolution wide column simulation v2} exhibits more complex behavior in its oscillations.

\autoref{fig:profile_snapshot_v0_crest} and \autoref{fig:profile_snapshot_v0_trough} show the interior structure of the column at the maximum and minimum heights of the oscillation, respectively.  The \lz{fact that the sides and the center are not always moving in the same direction through the course of the oscillation} is caused by the sides having greater cooling efficiency when the entire column is at its maximum height.  They therefore lose vertical pressure support first and collapse faster, thereby exposing \lz{more central} regions which can then cool more efficiently and collapse.  The sides also rebound first after reaching minimum height, covering up the still collapsing center and irradiating it with more photons (second
and fourth panels of \autoref{fig:profile_snapshot_v0_trough}).  This ultimately provides extra pressure support to the center causing it to \lz{expand vertically again} so that the entire column reforms a single-peaked mound at maximum height.

\begin{figure*}
    \centering
	\includegraphics[width=0.8\textwidth]{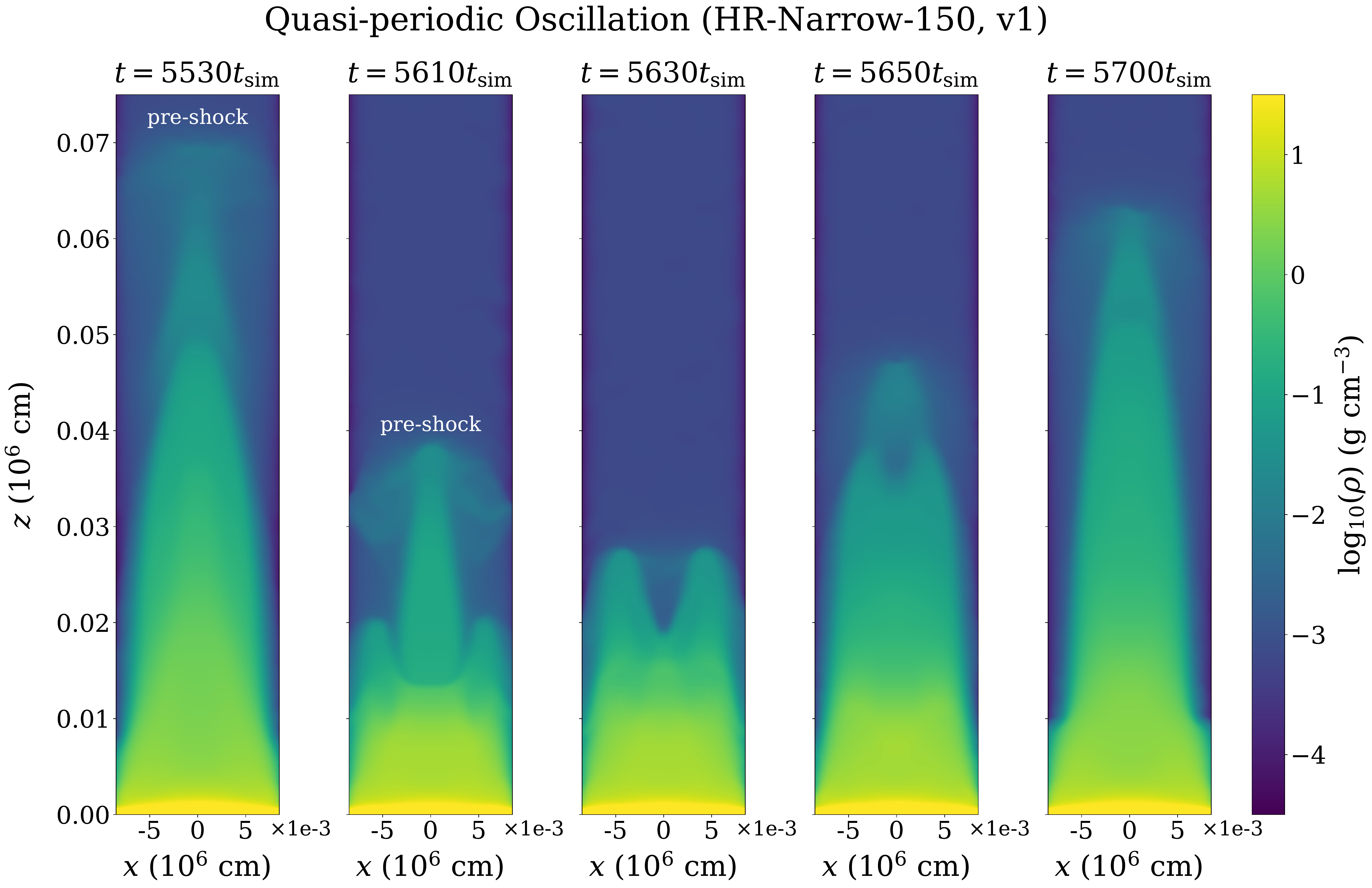}
    \caption{Selected 2D density snapshots \lz{over one full oscillation period $\simeq 170t_{\rm sim}=4.8\times10^{-5}$~s} in the narrow column simulation at higher accretion rate (v1) that are associated with the pre-shock accreting pattern.  \lz{2D profiles of other fluid quantities at the epochs of maximum (5530$t_{\rm sim}$) and minimum (5630$t_{\rm sim}$) vertical extent are shown in \autoref{fig:profile_snapshot_v1_crest} and \autoref{fig:profile_snapshot_v1_trough}, respectively.}
    }
    \label{fig:density_period_v1}
\end{figure*}

\begin{figure*}
    \centering
	\includegraphics[width=0.75\textwidth]{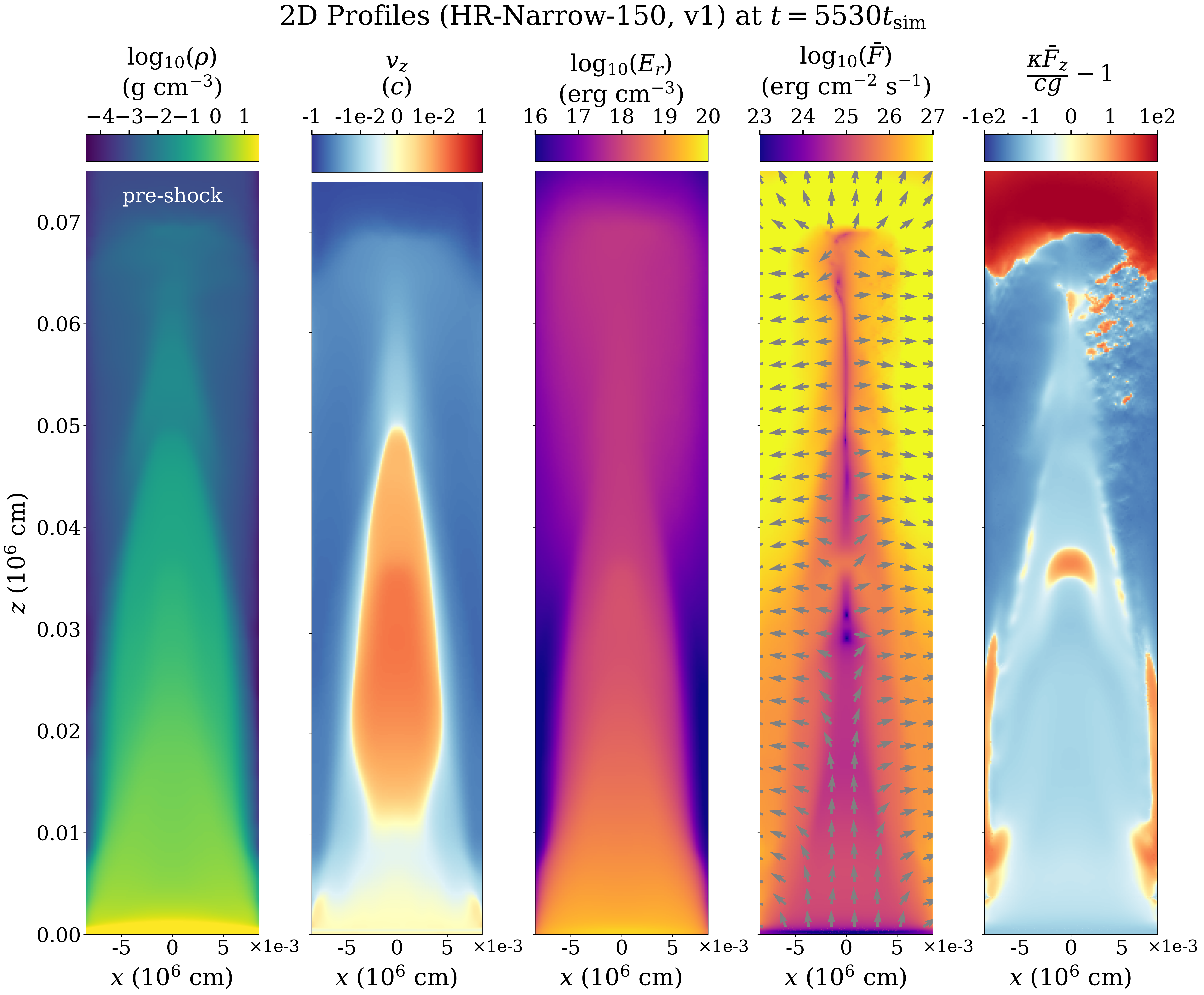}
    \caption{2D profiles at $t=5530t_{\mathrm{sim}}$ in the narrow column simulation at higher accretion rate (v1). Note that for the vector variables, the arrows represent the direction and the color bar indicates the magnitude.}
    \label{fig:profile_snapshot_v1_crest}
\end{figure*}

\begin{figure*}
    \centering
	\includegraphics[width=0.75\textwidth]{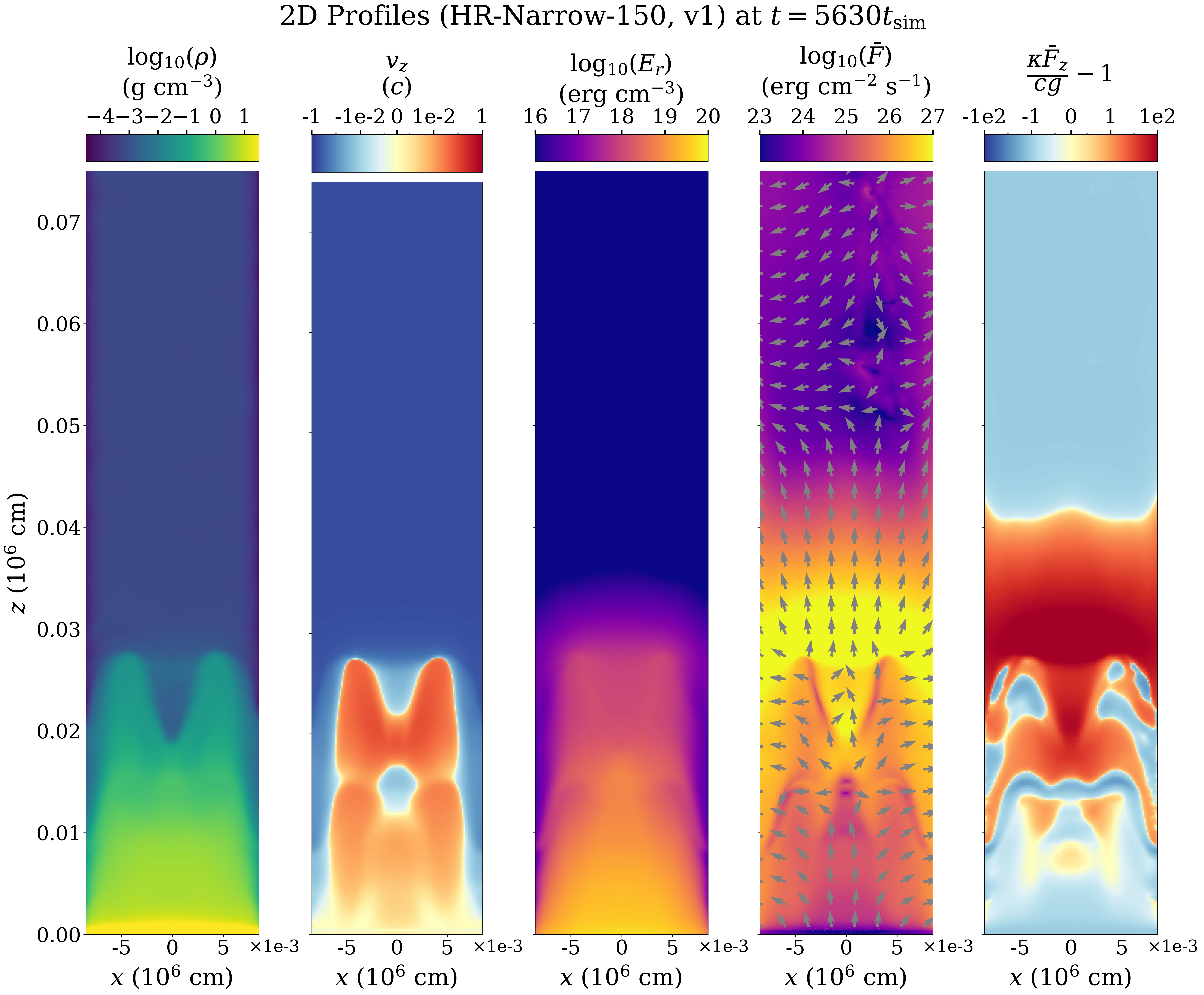}
    \caption{2D profiles at $t=5630t_{\mathrm{sim}}$ in the narrow column simulation at higher accretion rate (v1). Note that for the vector variables, the arrows represent the direction and the color bar indicates the magnitude.}
    \label{fig:profile_snapshot_v1_trough}
\end{figure*}

When we increase the accretion rate (v1) in the narrow accretion column, the overall oscillation in the light curve remains very similar (second
row of \autoref{fig:light_curve}).  \ob{\autoref{fig:density_period_v1} shows the density structure at different representative epochs during the oscillation. \autoref{fig:profile_snapshot_v1_crest} and \autoref{fig:profile_snapshot_v1_trough} show various quantities at the epoch of maximum and minimum height, respectively.  The} \lz{vertical} oscillation amplitude \lz{increases} with the increased accretion rate, although the overall oscillatory structure remains roughly the same. Note that in the first \lz{and second panels} of \autoref{fig:density_period_v1}, a pre-shock is quite prominent.  This happens on occasions when \lz{there is enhanced vertical escape of radiation which} can interact with the infalling material to form a shock above the main body of the sinking region. \lz{Indeed, we find that the radiation energy density and comoving frame radiation flux at altitude during these epochs are always larger than during epochs that lack a pre-shock.  The pre-shock formation} often happens at the peak of the radiative emission and thus prevents the accretion flow from reaching low altitude. Since the new shock forms at high altitude, the incoming energy can be dissipated in advance, causing a delay of the energy supply to the low altitude region, which results in weaker oscillations in the next few periods that recharges the sinking region. This effect happens in both simulations v0 and v1, as shown in the top two panels of \autoref{fig:light_curve}, where the light curves exhibit a flat-topped shaped oscillation which are sometimes followed by several narrower peaks, or even quiescence. We have confirmed that all flat-topped oscillations in both simulations are associated with a pre-shock accreting pattern (as in the first and second panels of \autoref{fig:density_period_v1}), and peak-shaped oscillations which lack this pre-shock structure (as in the first panel of \autoref{fig:density_period_v0}).  Flat-topped oscillation epochs with pre-shocks are more common in the higher accretion rate simulation v1.

\begin{figure*}
    \centering
	\includegraphics[width=\textwidth]{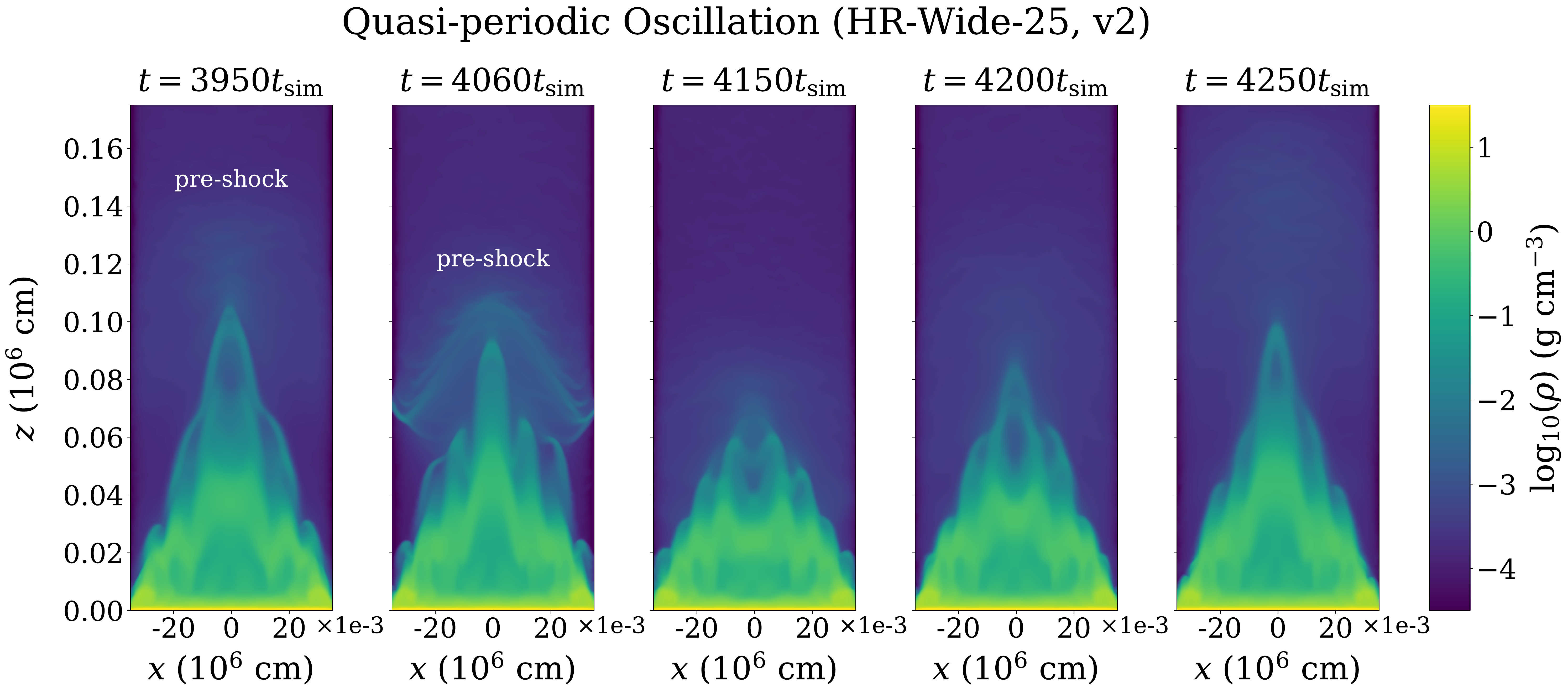}
    \caption{Selected 2D density snapshots \lz{over one full oscillation period $\simeq 300t_{\rm sim}=8.4\times10^{-5}$~s} in the high-resolution wide column simulation (v2).  \lz{In contrast to the narrow column simulations v0 (\autoref{fig:density_period_v0}) and v1 (\autoref{fig:density_period_v1}), this simulation shows complex multiple peaks at all phases of the oscillation.}  \lz{2D profiles of other fluid quantities at the epochs of maximum (3950$t_{\rm sim}$) and minimum (4150$t_{\rm sim}$) vertical extent are shown in \autoref{fig:profile_snapshot_v2_crest} and \autoref{fig:profile_snapshot_v2_trough}, respectively.}
    }
    \label{fig:density_period_v2}
\end{figure*}

\begin{figure*}
    \centering
	\includegraphics[width=0.975\textwidth]{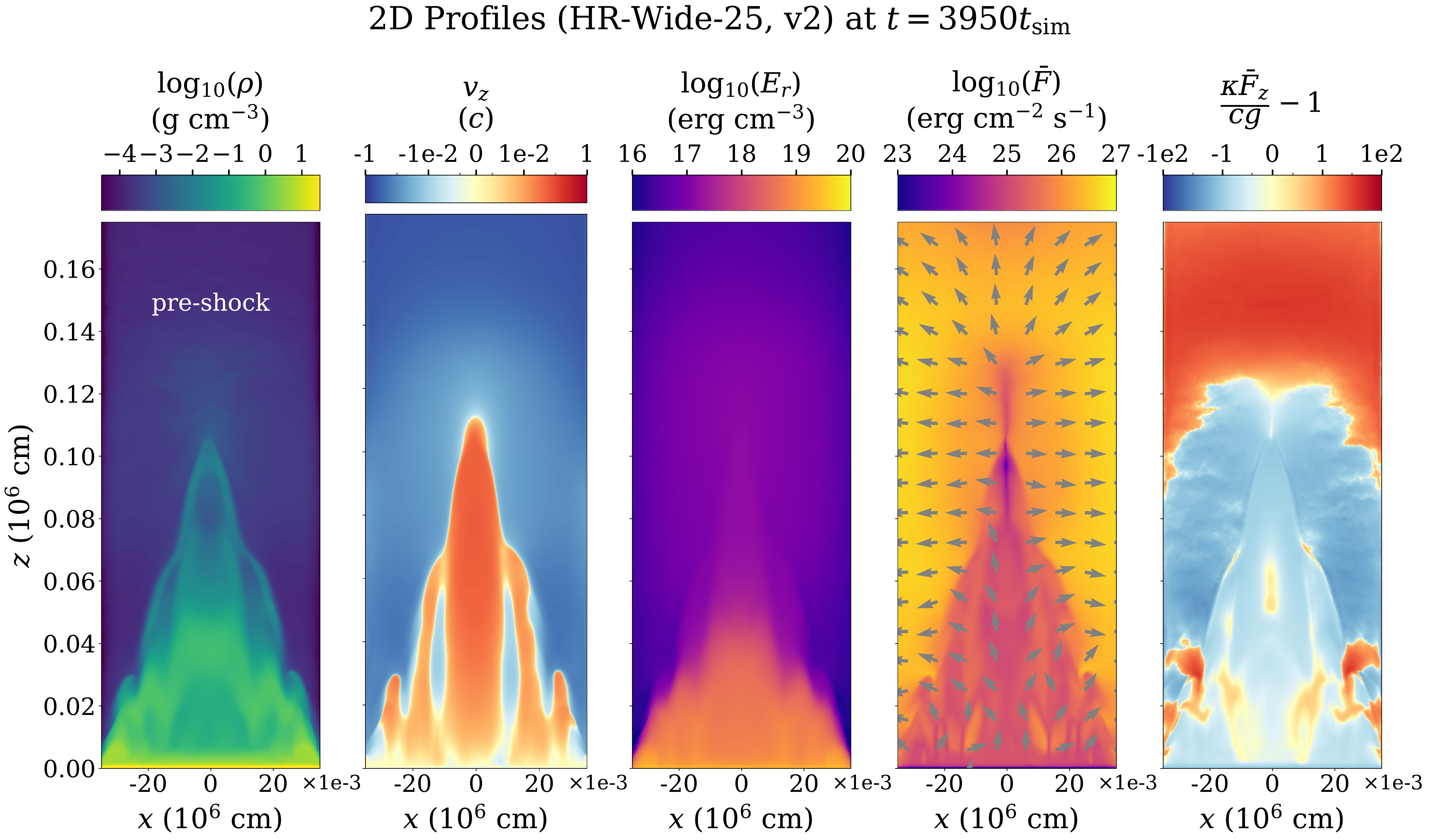}
    \caption{2D profiles at $t=3950t_{\mathrm{sim}}$ in the high-resolution wide column simulation (v2). Note that for the vector variables, the arrows represent the direction and the color bar indicates the magnitude.}
    \label{fig:profile_snapshot_v2_crest}
\end{figure*}

\begin{figure*}
    \centering
	\includegraphics[width=0.975\textwidth]{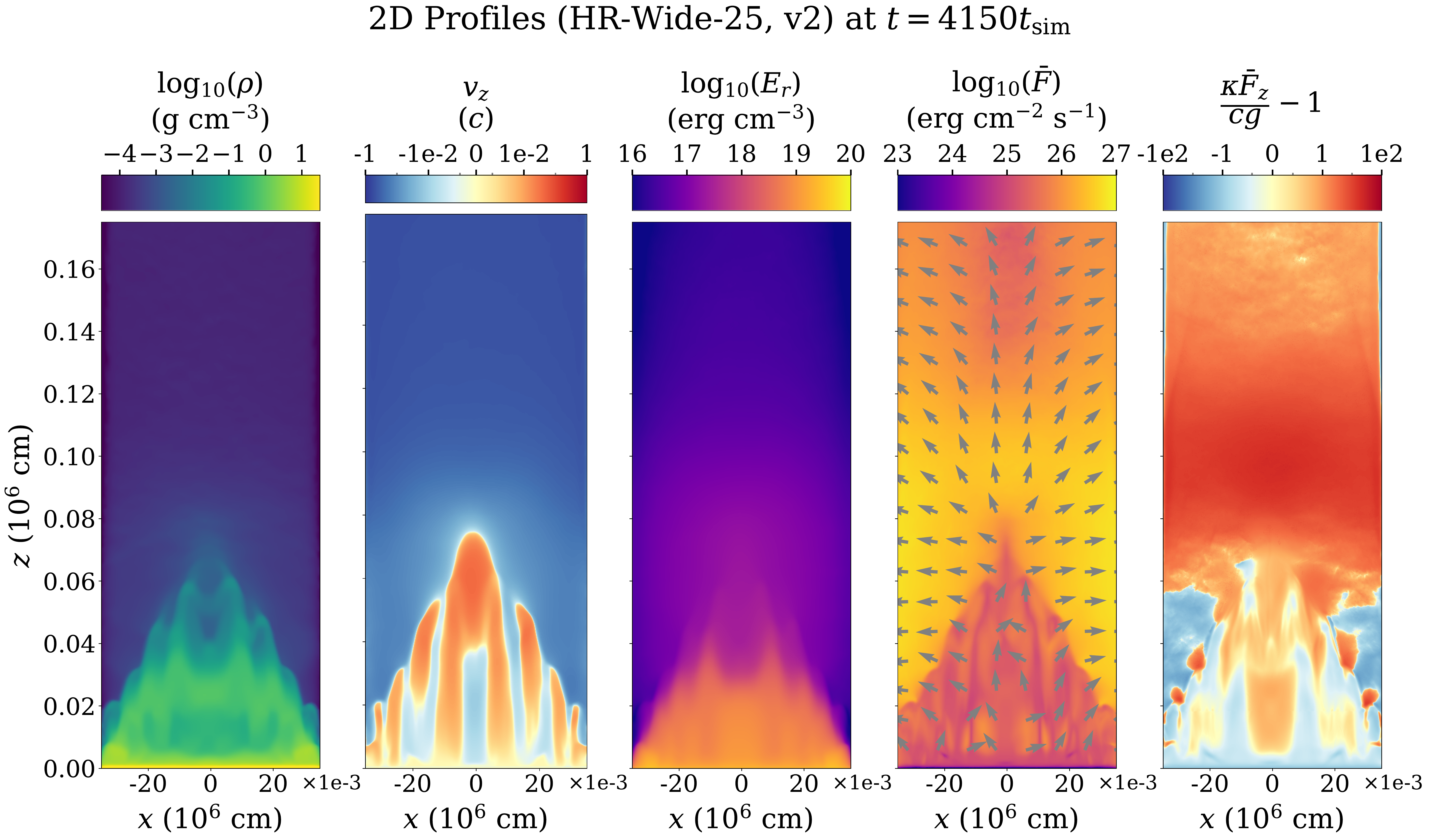}
    \caption{2D profiles at $t=4150t_{\mathrm{sim}}$ in the high-resolution wide column simulation (v2). Note that for the vector variables, the arrows represent the direction and the color bar indicates the magnitude.}
    \label{fig:profile_snapshot_v2_trough}
\end{figure*}

\subsubsection{Oscillatory behavior in wide accretion column}
\label{sec:osc_wide_column}
\lz{In simulation v2, we increase the column width with respect to the fiducial simulation v0, maintaining the same total accretion rate and therefore reducing the local Eddington ratio.}  \ob{\autoref{fig:density_period_v2} shows the density structure at representative epochs during the oscillation.  \autoref{fig:profile_snapshot_v2_crest} and \autoref{fig:profile_snapshot_v2_trough} show quantities at the epochs of maximum and minimum height, respectively.}  \lz{In contrast to the transition between a single-peaked maximum vertical extent and a triple-peaked minimum vertical extent exhibited in the narrow column simulations v0 and v1, \autoref{fig:density_period_v2} shows that the wide column simulation v2 exhibits a more complex multi-peaked horizontal structure throughout the oscillation.}  \lz{The vertical channels in this multi-peaked structure allow more radiation to propagate upward and interact with the incoming accretion flow.  (The vast majority of the radiation still leaves from the sides of the domain, in agreement with the \BSmodel.)  This again causes the formation of a pre-shock \lz{near} the maximum vertical extent of the oscillation (first and second panels of \autoref{fig:density_period_v2}).  As in the narrow column simulations, this results in flat-topped shaped oscillations in the light curve which, however, are more prominent in v2 because the horizontal structure maintains the pre-shock over much of the oscillation time interval.  There are intervals in v1 where the light curve stays close to its quiescent value ($\sim 3100$ and $4100\times2.8\times10^{-7}$~s), and this also happens in v2 at $\sim2900\times2.8\times10^{-7}$~s.  This appears to be happening because the pre-shock dissipates and radiates much of the accretion power, which then is not stored in the column structure itself.  The amplitude of the oscillation in the wide column simulation v2 is larger than in the narrow column simulation, substantially in height and slightly in the luminosity variation.  Recall that simulations v0 and v2 have the same total accretion rate, but the wider column in v2 has a longer horizontal radiation diffusion time.}  This ultimately results in a taller, more massive column that accumulates a greater total radiation energy, and this in turn increases the oscillation time scale and the oscillation amplitude. 

Inside the sinking region \lz{in v2}, there is a curved density \lz{inversion at lower altitudes ($z\sim0.04\times10^6$~cm at the column center)}, which oscillates at much smaller amplitude compared to the shock front itself.  This nearly stationary structure differs from that usually assumed in 1D standard models of neutron star accretion columns in being a vertical density inversion within the sinking zone.  Such density \lz{inversions} can exist in the radiation dominant magnetized medium as long as the vertical radiation flux can fully provide the support by adjusting the radiation energy density gradient to scale proportionately to the gas density (see Section 2 and Appendix A in \citealt{Zhang2021} for details).  These \lz{density inversion} structures are \lz{(nearly) hydrostatic} entropy modes \citep{1992ApJ...388..561A, Zhang2021}.

\subsubsection{Porosity}
\label{sec:porosity}

Neutron star accretion columns can in principle emit super-Eddington fluxes through their sides because of the confinement by strong magnetic fields, and this is also happening here in our simulations.  In addition to this, density inhomogeneities can create an effectively porous medium that permits an overall vertical super-Eddington flux without creating an overall radiation pressure force that exceeds gravity, simply because the flux will tend to be larger in the lower density regions (e.g. \citealt{1998ApJ...494L.193S, 2001ApJ...551..897B}).  Our accretion column simulations clearly exhibit substantial density inhomogeneities, and we
have attempted to quantify this by defining
\begin{equation}
    {\cal P}(z,t)\equiv\frac{\left<\rho\kappa\right> \left<|\bar{F}_{z}|\right>}
                 {\left<\rho\kappa|\bar{F}_{z}|\right>}
                 \quad, 
\end{equation}
as a height and time-dependent porosity factor with respect to vertical radiation transport, where the angle brackets refer to a horizontal average.  Porosity factors greater than unity indicate that the average fluid-frame vertical flux $\bar{F}_{z}$ is producing less horizontally-averaged vertical radiation pressure than would be the case for a horizontally homogeneous medium.  Depending on the epoch, we sometimes measure porosity factors as high as ten at certain heights, but this is generally due to the shape of the surface of the accretion column, e.g. significant vertical flux exists in the low density side regions of the mound-shaped column, or we get a notch of low density at the top of the center of the column as in the middle panels of \autoref{fig:density_period_v0} and \autoref{fig:density_period_v1}.  More \lz{relevant to the support of the column itself} is that we find
that the somewhat static low density region below $z\sim0.025\times10^6$~cm in the wide column simulation v2 (see \autoref{fig:density_period_v2}) has a porosity of $\simeq 3$ at all epochs.

\subsection{Time-averaged profiles}
\label{sec:profile_time_ave}

\begin{figure*}
    \centering
	\includegraphics[width=0.86\textwidth]{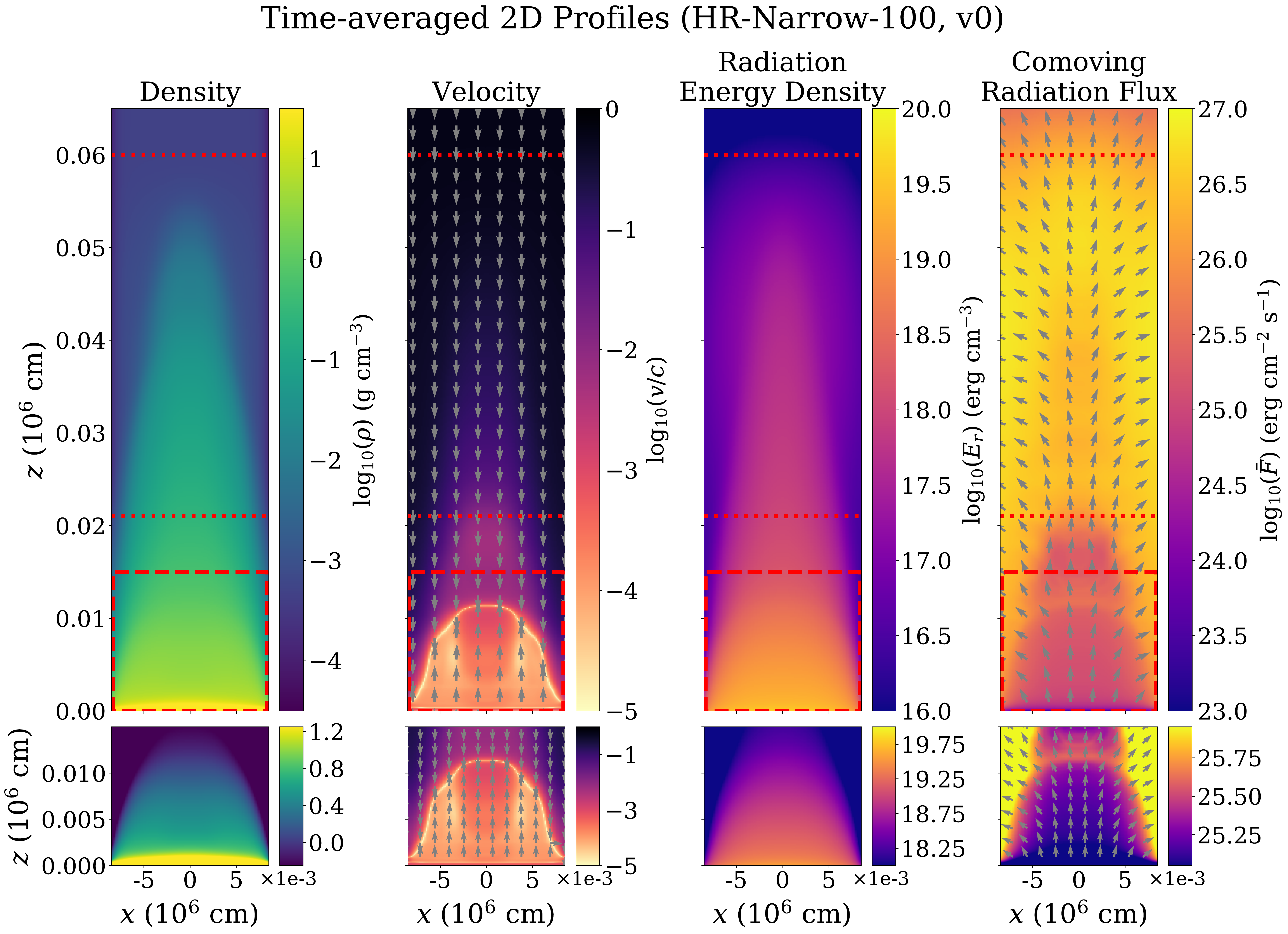}
    \caption{2D time-averaged profiles of the fiducial simulation (v0) after the simulation finishes its relaxation roughly at $t=2000t_{\rm sim}$. Note that for the vector variables, the arrows just represent the direction. \lz{The horizontal red dotted lines delineate the region affected by the vertically oscillating accretion shock, and are also shown in the 1D profiles in \autoref{fig:time_avg_1D}}.  The lower panels show the regions indicated by the red dashed boxes in the upper panels, but with different color bar scales to show more detail.}
    \label{fig:time_avg_2D_v0}
\end{figure*}

\begin{figure*}
    \centering
	\includegraphics[width=0.86\textwidth]{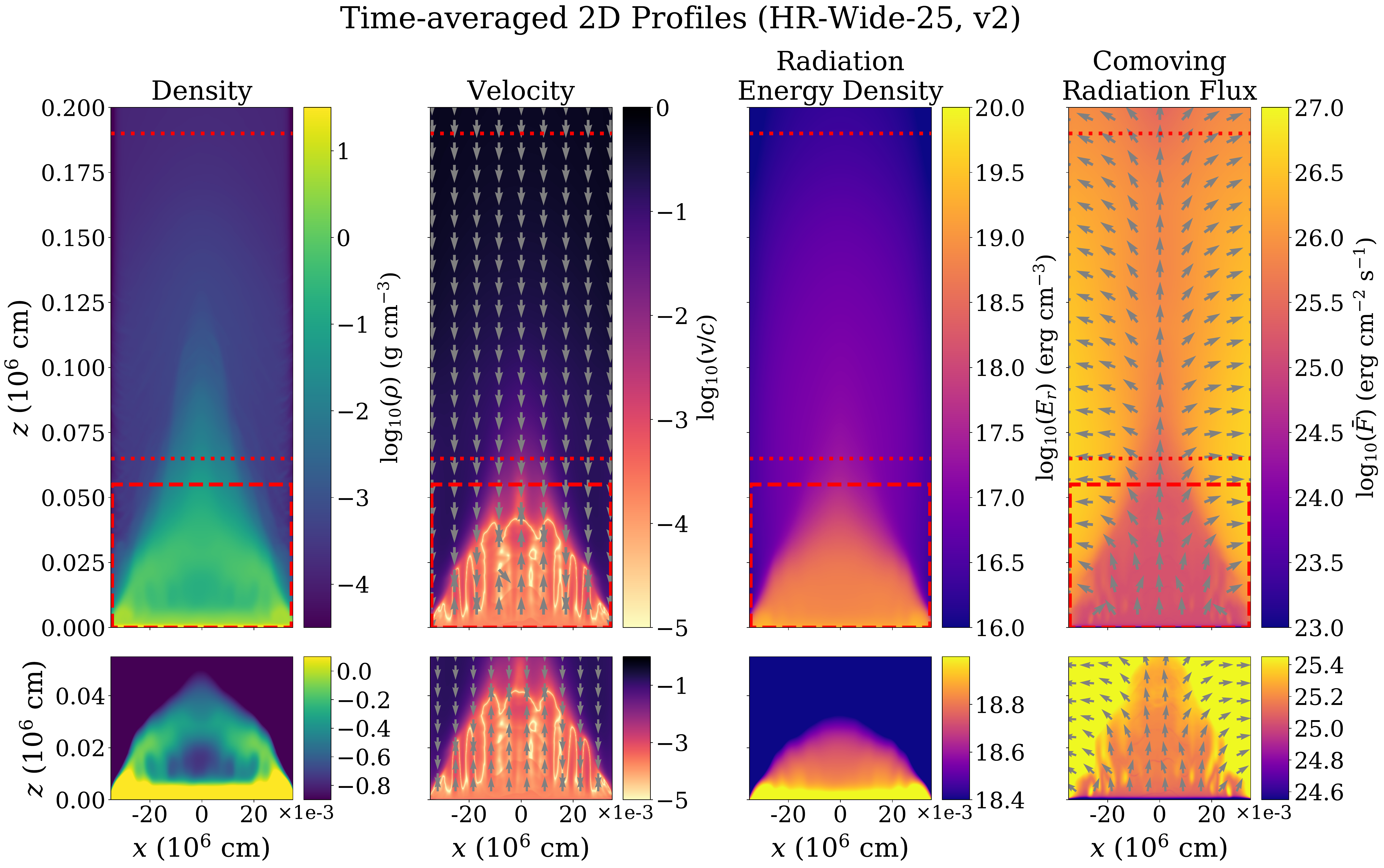}
    \caption{Same as \autoref{fig:time_avg_2D_v0}, but for the high-resolution wide column simulation (v2).}
    \label{fig:time_avg_2D_v2}
\end{figure*}

\begin{figure*}
    \centering
	\includegraphics[width=0.8\textwidth]{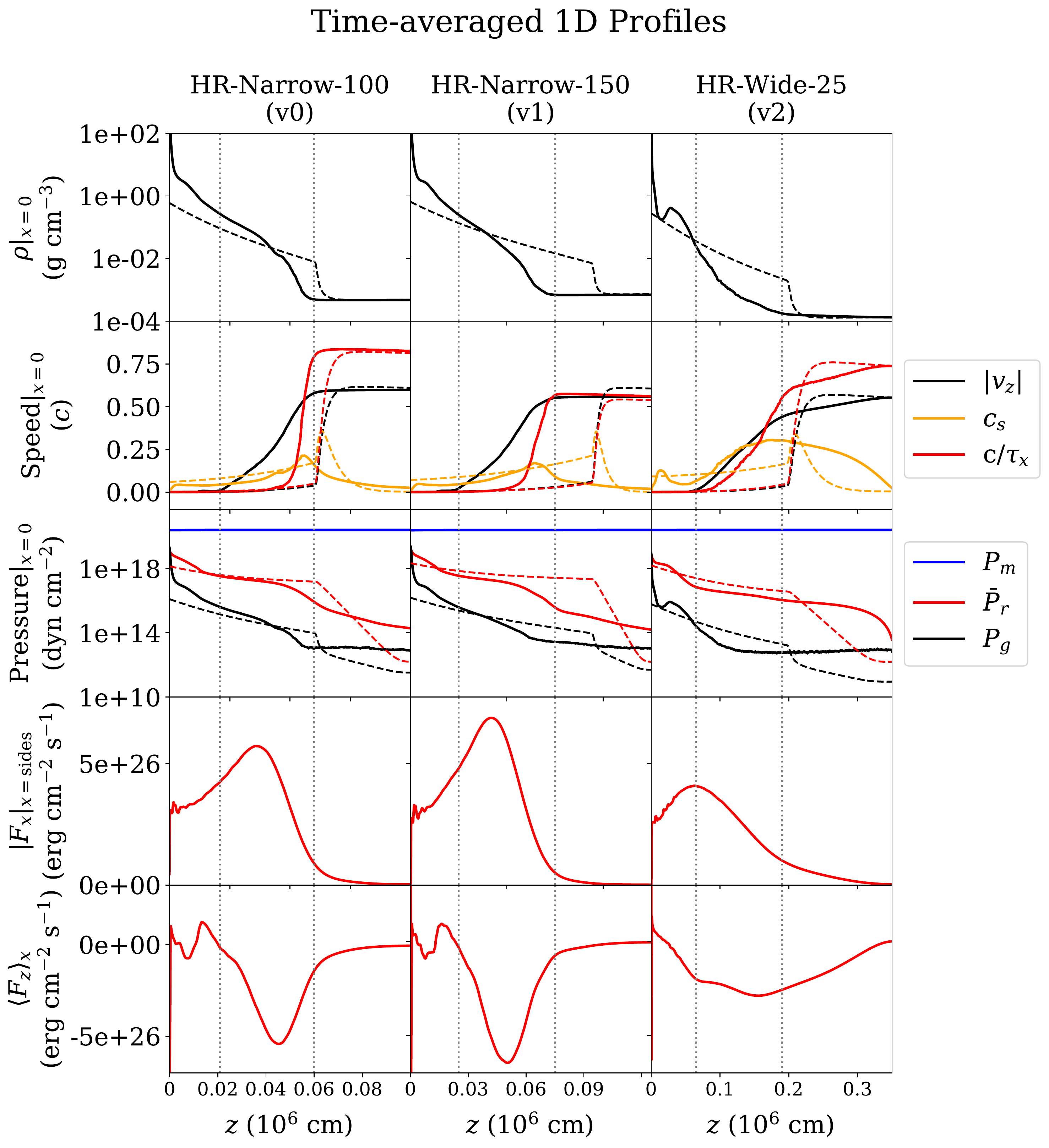}
    \caption{Time-averaged vertical profiles of density (top row), various speeds (second row), pressures (third row) at the $x=0$ center of the column from our three high resolution simulations as labelled in the \lz{column titles}. Also shown are the horizontal radiation flux averaged over both sides (fourth row) and the vertical radiation flux horizontally averaged over the column (last row).  Dashed curves in these figures show the initial condition, which also represents the expectations from the 1D standard model.  Vertical dotted lines indicate the \lz{vertical} spatial extent of the center of the column \lz{($x=0$)} \lz{over which the oscillating shock moves.}}
    \label{fig:time_avg_1D}
\end{figure*}

The high-frequency oscillation is challenging to observe with existing X-ray facilities because it requires the instrument to have enough signal to noise over these short time scales $\sim10~\mathrm{\mu s}$ of oscillations that are not perfectly coherent.  Partly for this reason, and partly to compare with the 1D standard model, we now discuss the time-averaged structure of our simulated accretion columns.  \lz{This is shown in \autoref{fig:time_avg_2D_v0} for the narrow column simulation v0 and in \autoref{fig:time_avg_2D_v2} for the wide column simulation v2.}

\lz{More detail of the time-averaged structure of all three of our high resolution simulations is shown in \autoref{fig:time_avg_1D}.  This depicts the} vertical profiles of various quantities all measured at the $x=0$ center of the column, as well as the horizontal radiation flux averaged over the two sides of the column and the horizontally averaged vertical radiation flux.  Also plotted in this figure are the initial conditions (dashed lines) for each of the three simulations, which represent the structure predicted by the 1D standard model.  The major differences between this model and the time-averaged profiles of the simulations come from the nonlinear oscillatory behaviors in the accretion column. The discontinuity of the 1D profile calculated from the 1D standard model is the location of the static accretion shock front. These discontinuities are smoothed out in the time-averaged profiles of our simulations because of the time-averaging over the oscillations.  The vertical extent of the region affected by the oscillating shock in the time-averaged profile is indicated by vertical dotted lines in \autoref{fig:time_avg_1D}.
The top and bottom of this region roughly correspond to the following locations in the time-averaged vertical velocity profile:

\begin{enumerate}
    \item The base of the shock oscillation region roughly corresponds to where the vertical flow velocity transitions from being dynamical in magnitude to the much smaller settling speeds within the low altitude regions of the column.  \lz{Below this height, the vertical velocity is always the small settling speed, but above this height, the time-averaged velocity is much larger because it averages the free-fall speed of material above the shock when it is low to the settling speed below the shock when it is high.}
    \item The top of the shock oscillation region corresponds to the point where the velocity of vertical flow first deviates from (the radiatively decelerated) free-fall speed.  It also corresponds to where the time-averaged density first starts to rise inward, although there is no sharp density discontinuity as it has been smeared out in the time-average.
\end{enumerate}

Most of the radiation escapes from the sides of our oscillating columns, and this of course remains true in the time-average.  The second to last row of \autoref{fig:time_avg_1D} shows the vertical profile of this sideways emission in each of our three high resolution simulations, and it is apparent that the sideways emission is dominated by the regions below the maximum height of the shock.  All of this is consistent with the 1D standard model. However, the time-averaged vertical radiation fluxes in the lab frame are mostly negative as shown in the bottom row of \autoref{fig:time_avg_1D}. Note that the vertical radiation flux in the fluid frame must remain positive to support the column structure so the accretion power must be injected into the sinking region via advection associated with the oscillatory motion.

As highlighted by the red dashed box in the first column of \autoref{fig:time_avg_2D_v2}, the 2D time-averaged density profile of the wide column simulation still exhibits a \lz{density inversion below the shock oscillation region.  (This is also evident in the upper right panel of \autoref{fig:time_avg_1D}.)} In other words, the smaller amplitude oscillating density \lz{inversion} below the instantaneous shock location discussed in the last paragraph of \autoref{sec:osc_wide_column} has not been time-averaged out.  This confirms that it is relatively steady even in the oscillatory column.  In the 2D time-averaged velocity profiles (second columns of both \autoref{fig:time_avg_2D_v0} and \autoref{fig:time_avg_2D_v2}), significant small-scale horizontal structure only exists in the wide column simulation, and this structure is also visible in all the other profiles (density, radiation energy density, and co-moving radiation flux).  These structures all have a characteristic horizontal length scale, and this is the same scale that is apparent in the instantaneous oscillatory behavior.  The fact that this has not been smoothed out in the time average suggests that this is a robust scale due to some underlying physics.  In \autoref{sec:resolution_dependence}, we suggest that this is due to photon bubble instability.

\subsection{Physical Origin of Accretion Column Oscillations}
\label{sec:interpretation_oscillation}

\begin{figure*}
    \centering
	\includegraphics[width=1.0\textwidth]{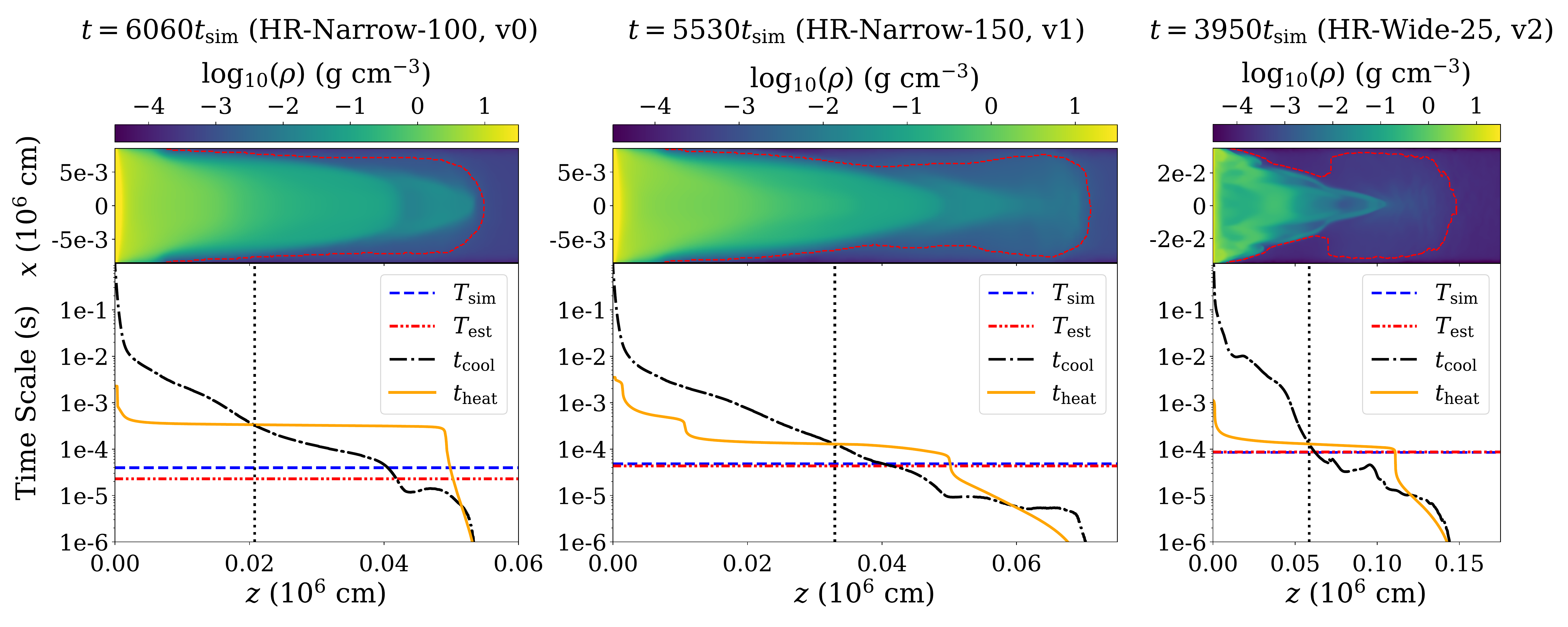}
    \caption{\lz{Comparison of different time scales at
    particular epochs corresponding to the maximum vertical extent of the column oscillation in simulations v0, v1, and v2 from left to right as indicated.  The 2D colour plots show the instantaneous density distribution at these epochs, with the red dashed contour indicating a density corresponding to the post-accretion shock density in the middle of the column ($x=0$). The black dot-dashed curve in each plot indicates the horizontal cooling time through radiative diffusion at each height.  The orange solid curve is the time it takes heat to diffuse or advect vertically from the top of the column down to the height indicated.  The vertical dotted lines indicate the location where these two time scales are equal, and are consistent with the location of the accretion shock at epochs of minimum vertical extent in the oscillation in the simulations.  The red dot-dot-dashed lines show our estimate of the oscillation period from equation \autoref{eq:T_est}, which agrees well with the measured oscillation periods in the simulations (blue dashed lines).} }
    \label{fig:freq_est_all}
\end{figure*}

\lz{
The oscillation periods in the light curves shown in \autoref{fig:light_curve} are ($\simeq4-8\times 10^{-5}\ \mathrm{s}$), which is comparable but longer than the free-fall timescale ($\sim 10^{-5}\ \mathrm{s}$).
We contend that the physical origin of these oscillations is not dynamical, but thermal.  It is fundamentally due to the inability of vertical advection in the settling flow and radiative diffusion to spread the accretion power liberated mostly below the accretion shock to the base of the column on a time scale short enough to balance the radiative cooling from the sides of the column.  We demonstrate this quantitatively here with an approximate analysis of the heat transport and cooling time scales.  Along the way, we find that we can also perform a good quantitative estimate of the oscillation period. }

\lz{We begin by considering epochs in the oscillations when the columns have reached their maximum vertical extent:  $t=6060t_{\rm sim}$ for v0, $t=5530t_{\rm sim}$ for v1, and $t=3950t_{\rm sim}$ for v2.  Because the oscillation period is longer than the free fall time, the column at these epochs is still in approximate vertical hydrostatic equilibrium.  We can then estimate how long it takes heat to be transported from the accretion shock at height $z_{\rm sh}$ in the middle ($x=0$) of the column down to a given height $z$:
\begin{equation}
      t_{\mathrm{heat}}(z) = \frac{z_{\rm sh} - z}{v_{\mathrm{heat}}}
      \quad.
\end{equation}
Here $v_{\mathrm{heat},z}$ is the heat transport speed at height $z$, given by
\begin{equation}
        v_{\mathrm{heat}} = \frac{-F_{z}}{\bar{E}_{r}}\Bigg|_{x=0}
        \quad,
\end{equation}
where $F_{z}$ is the lab frame radiative flux and $\bar{E}_r$ is the fluid frame radiation energy density.}

\lz{To estimate the cooling time, it is important to account for the mound shape of the accretion shock which bounds the mound.  We use the shock front gas density at $x=0$ to roughly contour the mound shape, as shown by the red dashed contours in the 2D density profiles of \autoref{fig:freq_est_all}. We can then compute the timescale of horizontal radiation diffusion ($t_{\mathrm{cool}}$) as
\begin{equation}
    t_{\mathrm{cool}}(z) = \frac{(x_r - x_l)/2}{c/\tau_x}
\end{equation}
where $x_l$ and $x_r$ refer to the left and right boundary of the contoured sinking region, respectively. The horizontal optical depth within the sinking region is defined by 
\begin{equation}
    \tau_x = \frac{1}{2}\int_{x_l}^{x_r}\rho \kappa dx 
\end{equation}
}

\lz{ As shown in \autoref{fig:freq_est_all}, $t_{\rm cool}<t_{\rm heat}$ in the upper regions, indicating that heat transport from the accretion shock cannot balance the cooling from the sides.  The accretion shock must
therefore collapse, quasi-hydrostatically, down to the depth where $t_{\rm cool}\simeq t_{\rm heat}$.  The vertical dotted line shows the location where the two time scales are equal, and below this line $t_{\rm cool}>t_{\rm heat}$ so that heat from the shock will exceed cooling.  The column will therefore become over-pressured, and drive the shock back upward.  We therefore expect the shock to oscillate up and down, with its lowest height being approximately the location of the vertical dotted line in \autoref{fig:freq_est_all} where these two time scales are in balance (hereafter $z_{\rm bot}$):  0.021$R_{\star}$ for v0, 0.033$R_{\star}$ for v1, and 0.059$R_{\star}$ for v2.  In fact, these heights are in excellent agreement with the lowest shock heights that we measure from the simulations (see $t=6150t_{\rm sim}$ for v0 in \autoref{fig:density_period_v0}, $t=5630t_{\rm sim}$ for v1 in \autoref{fig:density_period_v1}, and $t=4150t_{\rm sim}$ for v2 in \autoref{fig:density_period_v2}). }

\lz{The physics of this quasi-hydrostatic oscillation is therefore fundamentally related to heat transport and cooling, and we can further demonstrate this by estimating its period from a measure of the global cooling time of the upper portion of the column that is participating in the oscillation.  We do this by computing the enclosed radiation energy ($\Delta E_r$) and comparing it to the radiative heating ($F_{\mathrm{heat}}$) from the top ($z_{\rm sh}$) and the bottom ($z_{\rm bot}$). (The bottom heating amounts to at most 28\% of the total.)  Since the radiative cooling mainly depends on the side area, which varies considerably during the course of the oscillation, we adopt the sideways time-averaged radiation flux to compute the radiative cooling ($F_{\mathrm{cool}}$). Our estimate of the oscillation period ($T_{\mathrm{est}}$) then proceeds as
\begin{subequations}
\begin{align}
    \Delta E_r &= \int_{z_{\mathrm{bot}}}^{z_{\mathrm{sh}}} \int_{x_l}^{x_r} E_r(t_{\rm peak},x,z) dx dz
    \quad, 
    \\
    W &= \int_{z_{\mathrm{bot}}}^{z_{\mathrm{sh}}} \int_{x_l}^{x_r} \bar{P}_r(t_{\rm peak},x,z) \frac{\partial v_z (t_{\rm peak},x,z)}{\partial z} dx dz
    \quad, 
    \\
    F_{\mathrm{heat}} &= \int_{x_l}^{x_r} \left[F_{z}(t_{\rm peak},x,z_{\mathrm{bot}}) + F_{z}(t_{\rm peak},x,z_{\mathrm{sh}}) \right] dx
    \quad, 
    \\
    F_{\mathrm{cool}} &= \int_{z_{\mathrm{bot}}}^{z_{\mathrm{sh}}} \left[ \left< F_{x}(t,x_l,z) \right>_t + \left< F_{x}(t,x_r,z) \right>_t \right] dz
    \quad, 
    \\
    T_{\mathrm{est}} &= \frac{2\Delta E_r}{F_{\mathrm{cool}} - (F_{\mathrm{heat}} + W)}
    \quad, 
    \label{eq:T_est}
\end{align}
\end{subequations}
\lz{where $\bar{P}_r$ is the fluid-frame radiation pressure and $v_z$ is the gas vertical velocity, \lz{and $t_{\rm peak}$ refers to when the oscillating column is at its maximum extent. } The $pdV$ work $W$ is volume-integrated within the contoured sinking region.}  The factor of 2 in \autoref{eq:T_est} is based on the approximation that the same amount of time is spent in the collapse and expansion phases, which is definitely not true due to the nonlinear nature of this oscillation. Nevertheless, as shown in \autoref{fig:freq_est_all}, $T_{\mathrm{est}}$ is in excellent agreement with the direct measurements ($T_{\mathrm{sim}}$) of the oscillation period from the simulations, particularly in v1 and v2.}  \lz{Note that among v0, v1, and v2, the $pdV$ work $W$ is only $\sim$1\% of $F_{\rm heat}$. The distributed adiabatic compression is therefore negligible and cannot support the column.}

\subsection{Photon Bubbles and Resolution Dependence}
\label{sec:resolution_dependence}

\begin{figure*}
    \centering
	\includegraphics[width=0.8\textwidth]{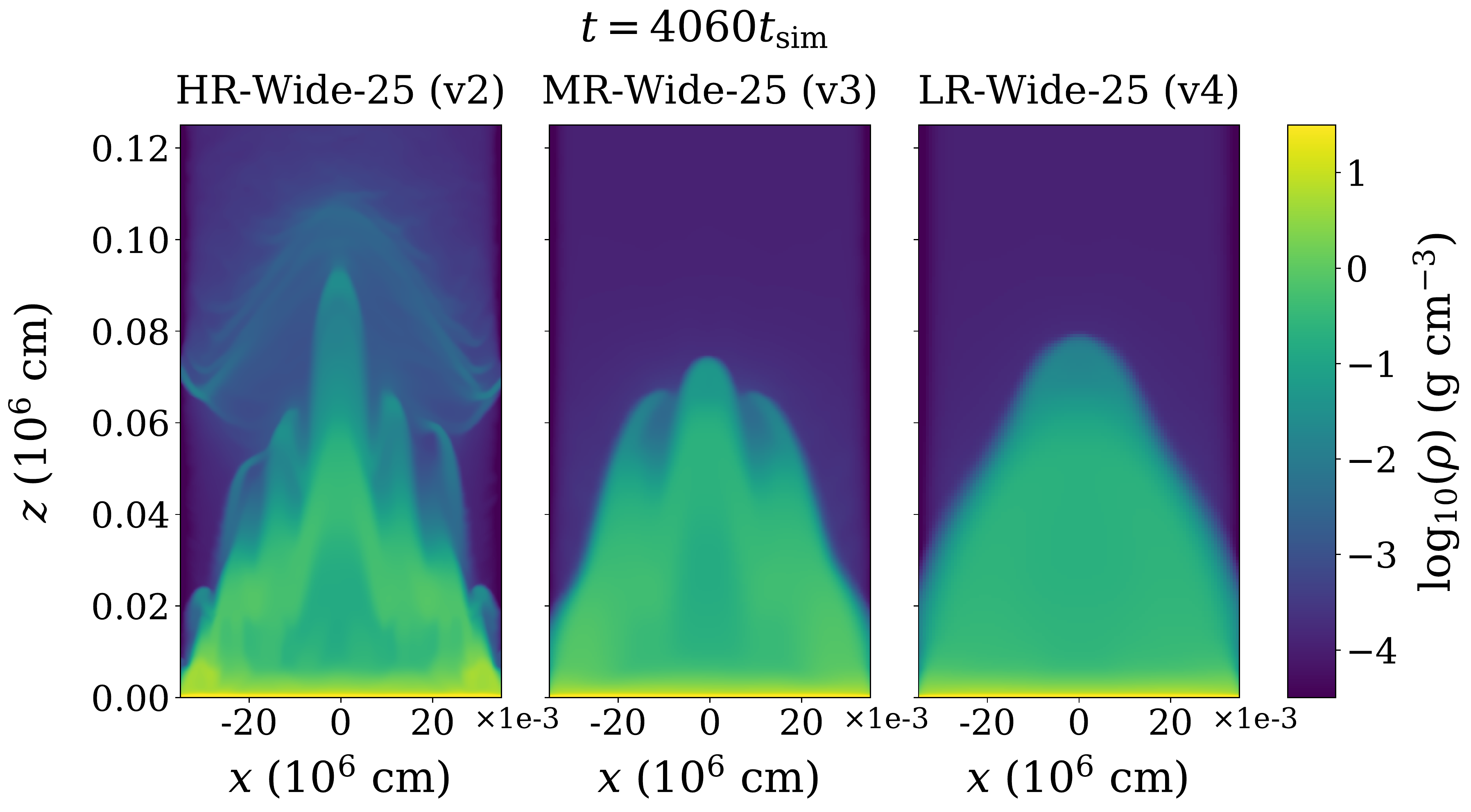}
    \caption{Snapshots of the density structure at $t=4060t_{\rm sim}$ in our wide column simulations at different resolutions:  high (left), medium (middle), and low (right).}
    \label{fig:horizontal_incoherence}
\end{figure*}

\subsubsection{\lz{Overview of Photon Bubble Instability}}
\lz{Ever since the pioneering linear instability analysis by \citet{1992ApJ...388..561A}, it has been expected that ``photon bubbles'' would be present in neutron star accretion columns.  The term ``photon bubbles'' suggests buoyant, bubble-like structures, but in fact this is not what this phenomenon represents, and the term is therefore highly misleading.  On length scales where photons diffuse rapidly, the ``photon bubble'' really amounts to a radiatively amplified acoustic wave that evolves to form a train of shocks
\citep{2001ApJ...551..897B,2005ApJ...624..267T}.  This rapid diffusion regime might manifest in the outer, low optical depth regions of neutron star accretion columns, but the cores of these columns are in the slow-diffusion regime.  There the instability is due to radiative amplification of entropy fluctuations, which is best understood in terms of phase lags between pressure and density perturbations caused by finite gas inertia, in response to radiative forcing.  We refer the reader to section 2 and Appendix A of \citet{Zhang2021} for a full description.  For a vertical magnetic field, the nonlinear outcome of the instability in this regime consists of nearly vertical wavefronts of density, with photons diffusing vertically more rapidly in the low density regions.  In a static atmosphere with no resupply of fresh material through accretion, this instability causes the atmosphere to collapse \citep{Zhang2021}.  The instability always grows fastest at shorter wavelengths, until either the radiation viscous length scale is reached \citep{Zhang2021}, or the scale height associated with gas pressure \citep{2003ApJ...596..509B}, whichever is larger.}

\lz{If the slow-diffusion version of this instability manifests itself in our simulations, we would expect nearly vertical wave fronts of density, i.e. vertical columnar patterns that move horizontally.  All three of our high resolution simulations v0, v1, and v2 manifest such structure, and the photon bubble instability is a likely candidate.  Unfortunately, demonstrating this in the narrow column simulations is difficult because the distinct columns only manifest during certain phases of the overall vertical oscillation, and moreover occur across a length scale that spans both the rapid and slow-diffusion regimes.  On the other hand, the wide column simulation has many high and low columnar structures in density, that are present at all times through the oscillation.  They also exhibit a distinct inward horizontal pattern speed that is evident in the animations.  We provide further evidence here that these structures are indeed an outcome of the photon bubble instability.}

\subsubsection{\lz{Resolution Dependence}}

\begin{figure*}
    \centering
	\includegraphics[width=\textwidth]{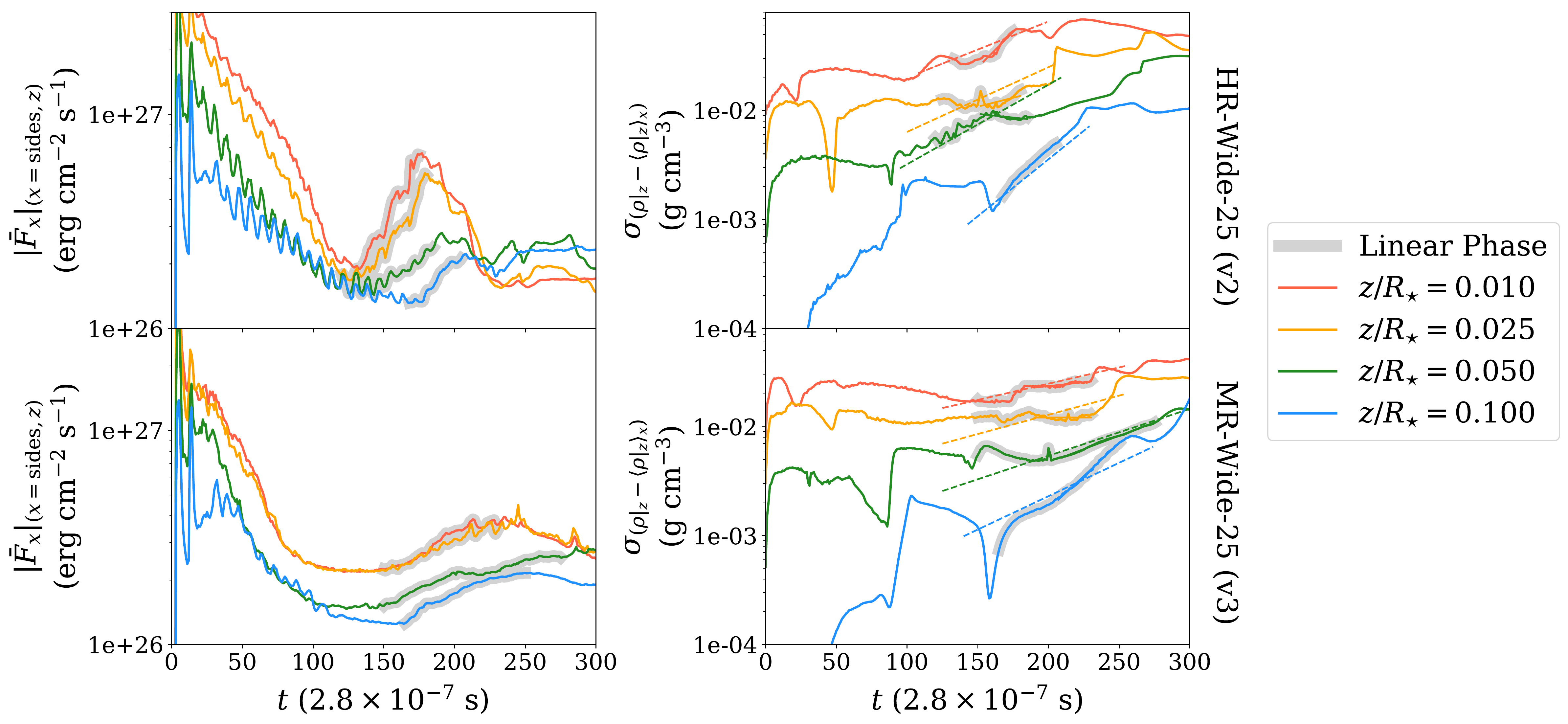}
    \caption{Linear growth of photon bubble instability in the relaxation epoch in the high-resolution and medium-resolution wide column simulations (v2 and v3). The left column shows the evolution of the fluid-frame radiation flux averaged over the two sides and the right column refers to the evolution of the density variation, where the linear growth phase is indicated by the grey shaded area.}
    \label{fig:linear_phase}
\end{figure*}

In \citet{Zhang2021}, we performed a resolution study on photon bubbles in static atmospheres that confirmed the behavior of increasing growth rate toward shorter wavelengths.  Here we conduct a similar resolution study of the wide column accretion simulation:  if the multi-peaked structure is due to photon bubble instability, the number of peaks should scale with spatial resolution.

We ran an additional two simulations v3 and v4 with factors of 2 and 4 lower resolution, respectively, compared to the high-resolution wide column simulation (v2). In \autoref{fig:horizontal_incoherence} we show snapshots of the density profile of these three wide column simulations at $t=4060t_{\mathrm{sim}}$. The dependence of horizontal structure with resolution is clear.  In the high-resolution simulation (v2), there are seven peaks in the shock fronts. In the medium-resolution simulation (v3) where the resolution is decreased by a factor of 2, there are only three shock front peaks.  In the low-resolution simulation (v4) where the resolution is decreased by another factor of 2, the shock structure is single-peaked.  The decreasing number of shock front peaks therefore tracks the decreasing factor of resolution. This resolution dependence is indicative of modes that grow fastest at the smallest resolvable length scales.  \lz{This resolution dependence is consistent with unstable modes that grow faster at shorter length scales, and is precisely the behavior of the linear photon bubble instability and its nonlinear outcome
\citep{Zhang2021}.}

\lz{None} of the high resolution simulations here, including the wide column simulation v2, resolve the viscous length scale, which is approximately 10 cm \citep{Zhang2021}, about a factor of 15 smaller than the grid scale in each simulation.  Resolving such small scales is very expensive and may not even be possible for more global simulations of neutron star accretion columns with existing computational resources. However, \lz{at least for the simulations here}, these horizontal structures only modify the column dynamics by increasing the oscillation amplitude without changing its period (see \autoref{fig:light_curve} which shows light curves for all our simulations).  Hence we have reached convergence in terms of characteristic time scales present in the light curves and the nonlinear dynamics.

\lz{Note that the light curve of the medium resolution simulation (v3) exhibits more noisy behavior than either the low resolution (v4) or high resolution (v2) simulations.  This is because v3 is at just high enough resolution to start exhibiting multi-peaked horizontal structure.  The light curve becomes better behaved with the addition of more multiple peaks at higher resolution.  That the multi-peaked structure in v2 produces enhanced vertical radiation transport can be seen by comparing this simulation to the lower resolution simulations v3 and v4.  Note that these latter simulations do not form a pre-shock (cf. \autoref{fig:horizontal_incoherence}).  As can be seen in the animations of v2, v3, and v4, both the radiation energy density and the flux at altitude are much higher in v2 compared to v3 and v4.  In fact, the vertical flux above $z=0.15\times10^6$~cm is at least ten times higher in v2 at all epochs.}

Although we believe that we have reached numerical convergence in terms of the light curve behavior, the presence of increasing horizontal structure with increased resolution implies that the time-averaged structures are not converged.  This may be an issue when comparing predictions of spectra and polarization from these time-averaged profiles with observations integrated over time scales longer than the very short oscillation period.  On the other hand, the \lz{additional density inversion that forms below the accretion shock oscillating region} appears to be robust, and is therefore likely to be independent of further increase in resolution.  Unfortunately, increasing resolution appears to exacerbate the variable inversion problem in our numerical algorithm and results in more noise in the gas temperature.  Clearly more work is needed to address this numerical problem.

\subsubsection{\lz{Linear Phase of Photon Bubbles}} 
\label{sec:pb_dynamics}

To further demonstrate that these horizontal structures are in fact related to photon bubbles, we searched for the linear phase of photon bubble growth and compared that behavior to the analytic linear theory  \citep{1992ApJ...388..561A, 1998MNRAS.297..929G, Zhang2021}.  Although the linear theory applies to static atmospheres \lz{with infinite horizontal extent}, not oscillating accretion columns, the radiation sound speed is still much greater than both the fluid velocity and radiation diffusion speed in the sinking region, so we are in the slow-diffusion regime with instantaneous near-hydrostatic equilibrium.  The photon bubble instability mechanism should therefore still operate, \lz{and because the photon bubble growth times are fast ($\sim10~\mathrm{\mu s}$), we should be able to observe their linear growth phase.}

Since the horizontal \lz{multi-peak structure only appears in the high-resolution (v2) and medium-resolution (v3) simulations, and not in the low resolution simulation (v4), we analyze just the first two.}  As shown in the left column of \autoref{fig:linear_phase}, immediately after launching the simulation, radiation leaves the column through the sides.  \lz{This flux declines until photon bubbles start to grow.  The linear phase is indicated by the grey shading and indicates approximately exponential growth, except perhaps at $z/R_\star=0.025$ (the orange curve) in density.  The dashed lines indicate the predicted local WKB maximum growth rates from the linear dispersion relation for photon bubbles in a {\it static, non-accreting} atmosphere \citep{Zhang2021}, which are in rough agreement with what we observe in the simulations. In particular, two facts are worth noting that are consistent with identification with photon bubbles:  the growth rates increase with height, and they increase with numerical resolution (v2 grows faster than v3).}

\section{Discussion}
\label{sec:discussion}

\subsection{Validity of the Parameter Regime in Simulations}
\label{sec:validity_param}
The typical magnetic fields inferred for accretion columns are $\gtrsim 10^{12}\ \rm G$. However, in our simulations, we cannot adopt such high field strengths because they exceed the tolerance of the variable inversion algorithm for resolving the gas pressure, which is crucial at the step of computing the radiative transport source term \autoref{eq:rad_source_term_fluid_frame}.  Therefore we have to maintain 
the ratio of gas pressure to magnetic pressure as large as possible in order to obtain sufficient precision in the numerical value of gas pressure. Fortunately, the main effect of the strong magnetic fields is to horizontally confine the gas, and there is very little effect on the vertical dynamics.  This allows us to adopt a magnetic field that is strong enough for gas confinement ($8 \times 10^{10}$ G in our case) and keep the correct dynamics of the accretion column.

One way in which the choice of magnetic field could be more significant is its effect on the electron scattering opacity.  We assume isotropic Thomson scattering, which is valid for the low field strength and high temperatures in our simulations here, and would continue to be valid up to field strengths approaching $\sim10^{12}$~G.  Magnetic scattering opacities are
likely to affect the nonlinear dynamics above these field strengths.

\subsection{Comparison with Previous Works}
The 1D models of accretion columns at high accretion rate by \citet{1976MNRAS.175..395B} and, more recently, \citet{2017ApJ...835..129W} neglect horizontal gradients in the column structure.  As such, these columns have a top-hat shape, i.e. they neglect the actual two-dimensional mound shape that must exist in reality.  Because of this assumed geometry, they necessarily under-estimate the horizontal \lz{radiation diffusion speed}, and therefore over-estimate the column height.  This explains in part why the time-averaged shock height from our simulations, is lower than the 1D model prediction of \citet{1976MNRAS.175..395B}, as shown in \autoref{fig:time_avg_1D}.
The other important difference between these 1D models and our simulations is that the simulations have dynamical, nonlinear oscillations.  In the stationary \BSmodel, accretion power can only be released in the region below the shock by the loss of gravitational potential energy by the slowly sinking material.  This suggests two problems: 1. when the sinking region has a small height, the gravitational energy liberated by the sinking gas is insufficient to support the column structure. 2. when the sinking region is high, the hydrostatic equilibrium can be hardly maintained by a process (vertical advection) that is necessarily much longer than the dynamical time.  In fact, these columns are not stationary, but undergo considerable dynamical motion up and down.  It is this dynamical motion that then redistributes accretion power vertically as the material radiates.  This dynamical mechanism of redistributing the accretion power liberated at the shock front is fundamentally why the oscillations exist and, indeed, persist.  The photon bubble instability itself \lz{is not} the cause of the oscillations.  \lz{However, photon bubbles enhance vertical diffusive transport and appear to} enhance the coherence of the light curve oscillation \lz{by aiding in the formation of pre-shocks that result in flat-topped luminosity variations in the light curve.} 

The pioneering numerical simulations of accretion columns by \citet{1989ESASP.296...89K} and \citet{1996ApJ...457L..85K} also exhibited oscillatory behavior with time scales between 0.1 and 1 ms. This is qualitatively consistent with our simulations, although their oscillation \lz{period} is slightly longer than ours ($\sim 0.05$ ms), likely due to the fact that our columns simulated here are smaller in height. The most recent numerical work on accretion columns prior to our work was done by \citet{2020PASJ...72...15K}, who also discovered the development of similar finger-shaped density structures. However, instead of injecting accreting material from the top boundary, \citet{2020PASJ...72...15K} initialized a uniformly distributed gas that free-falls from rest in the column domain, where the initial gas density was scaled according to the different accretion regimes.

Note that the simulations of \citet{1989ESASP.296...89K} and \citet{2020PASJ...72...15K} found that the multiple-bubble structure disappeared for lower accretion rates.  However, this may have been a resolution effect.  As the accretion rate decreases, the local Eddington ratio in the sinking zone also decreases, which leads to larger gas density because more gas pressure gradient is required to support the column. Therefore, the radiation diffuses more slowly in the low luminosity case. This directly shifts the peak of the photon bubble growth rate to shorter wavelengths (see Appendix A2 in \citealt{Zhang2021} for details) making it harder to resolve the instability.  Moreover, the geometric dilution introduced by any non-Cartesian geometry, e.g. spherical \citep{2020PASJ...72...15K} and dipolar \citep{1989ESASP.296...89K} coordinate systems, naturally decreases the resolution at higher altitude.  This loss of resolution will prevent the development of photon bubble behavior at short wavelengths.

\subsection{Observational Significance}
\label{sec:observ_app}
Our long-term plan is to quantify the dynamics and observables of neutron star accretion columns for X-ray pulsars above the critical luminosity \citep{2012A&A...544A.123B, 1976MNRAS.175..395B} through the use of numerical simulations. Because of our use of Cartesian geometry in this paper, we are forced to restrict consideration to \lz{modest} accretion rates and X-ray luminosities so that the column height is much less than the radius of the star.  We intend to pursue more global simulations of higher accretion rate, larger columns in future.

There are in fact some X-ray sources that pass through the accretion rate regime of our Cartesian simulations: high enough that the sinking zone is developed but also low enough that the height of the accretion shock front does not exceed $\sim R_{\star}$, in particular transient X-ray pulsars such as EXO2030+375, 4U0015+63, KS1947+300, and V0332+53 \citep{1989ApJ...338..373P, 1989ApJ...338..381W, 1990SvAL...16..345B, 1984ApJ...285L..15T, 2013A&A...551A...1R}. It is noteworthy that the emission patterns of these sources transition between pencil beam and fan beam, and this is usually interpreted in terms of the critical luminosity when an optically thick accretion column is thought to form, according to the \citet{1976MNRAS.175..395B} theory.  The existence of nonlinear oscillations and photon bubble dynamics, which we have shown here to result in a lower time-averaged height of the accretion column \lz{compared to} 1D models, probably does not affect this conclusion as an optically thick column is still required to host this dynamics.  Moreover, we have found here that photon bubble dynamics does not alter the overall fact that an accretion column tends to produce fan beam emission:  both the instantaneous and time-averaged cooling of the column is dominated by sideways emission from the column.

The main observable difference is the presence of high frequency variability ($\sim 20$~kHz), as also pointed out in the pioneering work of \citet{1996ApJ...457L..85K} who predicted significant variability power at frequencies $\sim$1-10~kHz.  Indeed, \lz{this has} been invoked to explain putative high frequency variability in the X-ray pulsar Cen X-3 (e.g. \citealt{2000ApJ...530..875J}).  However this \lz{observed} noise component may actually have been an artifact of the splitting up of photon counts in RXTE/PCA sub-bands \citep{2015MNRAS.451.4253R}. An attempt to analyze lightcurve data of the bright X-ray pulsar V0332+53 that avoided this problem was attempted by \citet{2015MNRAS.451.4253R}, who found no extra high frequency noise component to the level of 0.5 percent.  It would be worthwhile searching for high frequency features in the light curves of other X-ray pulsars given how robust are the oscillations that we have found here.  It is likely that the oscillation frequencies will depend on the column height and accretion geometry, and more global simulations that take into account the diverging magnetic field geometry will be necessary to quantify this.

\section{Conclusions}
\label{sec:conclusions}
We summarize our conclusions as follows:
\begin{enumerate}
    \item Our simulations more or less agree with standard 1D models \lz{of supercritical accretion, in that an optically thick accretion column forms with an accretion shock at the top of the column.}  The column is supported against gravity by radiation pressure gradients, and most of the cooling takes place through the sides of the column (fan-beam radiation), as in the original models of \citet{Inoue1975} and \citet{1976MNRAS.175..395B}.  However, in order \lz{to establish thermal equilibrium, the sinking material must be supplied with heat at all altitudes to balance the local sideways cooling.}  Because \lz{the $PdV$ work on the sinking material is small, and the time scale of vertical heat transport from the accretion shock} is very much longer than the \lz{local cooling in the upper regions of the mound-shaped column, the column must cool and collapse.  This then produces a shorter, denser column with a longer sideways cooling time, which then overheats and re-expands.  This is the origin of the oscillations that we observe.}  \lz{Oscillations were also reported by \citealt{1996ApJ...457L..85K} and attributed to photon bubbles, but we suspect that they may have been due to the same mechanism that we have found here.}
    
    \item The shock fronts in our simulations in general have lower altitudes, both instantaneously and in the time-average, than those in the standard 1D model.  The basic reason for this is that the finite horizontal one-zone approximation of the 1D models under-estimates the efficiency of cooling from the sides due to the fact that the accretion column actually has more of a 2D mound shape.
    
    \item Photon bubbles do not appear to be \lz{directly} responsible for \lz{either the vertical oscillations or the enhanced cooling compared to 1D models.}  Instead, they are a further complication that produces a complex multi-peaked horizontal structure when the instability is spatially well-resolved. We have demonstrated that this is so by showing that these structures have similar linear growth rates and resolution dependencies to the linear photon bubble instability.  \lz{While we have not achieved numerical convergence on the photon bubble structures, the overall oscillation period of the column is not affected by spatial resolution, though the light curve has a smoother oscillatory shape at high spatial resolution.}
    
    \item Both the high-frequency oscillation and photon bubble dynamics facilitate the vertical redistribution of energy dissipated at the shock front to various altitudes so that the column structure can be maintained. Whether this continues to be the case for more global accretion columns in spherical/dipolar geometry will be the subject of our next paper.
\end{enumerate}

\section*{Acknowledgements}

\lz{We thank the anonymous referee for very thorough comments which greatly improved this paper.}  We also thank Mitch Begelman, Matt Middleton, and Jim Stone for useful conversations.  This work was supported in part by NASA Astrophysics Theory Program grant 80NSSC20K0525.  Resources supporting this work were provided by the NASA High-End Computing (HEC) Program through the NASA Advanced Supercomputing (NAS) Division at Ames Research Center.  We also used computational facilities purchased with funds from the National Science Foundation (CNS-1725797) and administered by the Center for Scientific Computing (CSC). The CSC is supported by the California NanoSystems Institute and the Materials Research Science and Engineering Center (MRSEC; NSF DMR 1720256) at UC Santa Barbara. The Center for Computational Astrophysics at the Flatiron Institute is supported by the Simons Foundation.

%%%%%%%%%%%%%%%%%%%%%%%%%%%%%%%%%%%%%%%%%%%%%%%%%%
\section*{Data Availability}

All the simulation data reported here is available upon request to the authors.


\bibliographystyle{mnras}
\bibliography{references} % if your bibtex file is called example.bib

%%%%%%%%%%%%%%%%%%%%%%%%%%%%%%%%%%%%%%%%%%%%%%%%%%

%%%%%%%%%%%%%%%%% APPENDICES %%%%%%%%%%%%%%%%%%%%%

% \appendix
% \section{Some extra material}
% If you want to present additional material which would interrupt the flow of the main paper, it can be placed in an Appendix which appears after the list of references.

%%%%%%%%%%%%%%%%%%%%%%%%%%%%%%%%%%%%%%%%%%%%%%%%%%

% Don't change these lines
\bsp	% typesetting comment
\label{lastpage}

\end{CJK*}
\end{document}